\documentclass[10pt,twocolumn,showpacs,preprintnumbers,amsmath,amssymb,aps,prl,superscriptaddress,longbibliography]{revtex4-2}
\usepackage{appendix}
\usepackage{bbm}
\usepackage{mathrsfs}
\usepackage{graphicx}
\usepackage{dcolumn}
\usepackage{bm}
\usepackage{braket}
\usepackage{amsmath}
\usepackage{amsfonts}
\usepackage[dvipsnames]{xcolor}
\usepackage[colorlinks=true,linkcolor=Blue,urlcolor=BlueViolet,citecolor=BlueViolet]{hyperref}
\usepackage{natbib}
\usepackage{multirow}

\begin{document}

\title{Data-Driven Learnability Transition of Measurement-Induced Entanglement}
\author{Dongheng Qian}
\affiliation{State Key Laboratory of Surface Physics and Department of Physics, Fudan University, Shanghai 200433, China}
\affiliation{Shanghai Research Center for Quantum Sciences, Shanghai 201315, China}
\author{Jing Wang}
\thanks{wjingphys@fudan.edu.cn}
\affiliation{State Key Laboratory of Surface Physics and Department of Physics, Fudan University, Shanghai 200433, China}
\affiliation{Shanghai Research Center for Quantum Sciences, Shanghai 201315, China}
\affiliation{Institute for Nanoelectronic Devices and Quantum Computing, Fudan University, Shanghai 200433, China}
\affiliation{Hefei National Laboratory, Hefei 230088, China}

\begin{abstract}
Measurement-induced entanglement (MIE) captures how local measurements generate long-range quantum correlations and drive dynamical phase transitions in many-body systems. Yet estimating MIE experimentally remains challenging: direct evaluation requires extensive post-selection over measurement outcomes, raising the question of whether MIE is accessible with only polynomial resources. We address this challenge by reframing MIE detection as a data-driven learning problem that assumes no prior knowledge of state preparation. Using measurement records alone, we train a neural network in a self-supervised manner to predict the uncertainty metric for MIE—the gap between upper and lower bounds of the average post-measurement bipartite entanglement. Applied to random circuits with one-dimensional all-to-all connectivity, our method reveals a learnability transition with increasing circuit depth: below a threshold the MIE can be effectively learned with resources that grow only polynomially with system size, whereas above it the required resources grow exponentially. This computational phase transition coincides with the breakdown of efficient classical simulation of the underlying quantum state. We further observe signatures of this transition on current noisy quantum devices. These results highlight the power of data-driven approaches for learning MIE and delineate the practical limits of its classical learnability.
\end{abstract}

%\date{\today}

\maketitle

Quantum entanglement is a central resource of quantum information science, enabling advantages in computation, communication, and sensing~\cite{horodecki2009, nielsen2010, wilde2017, degan2017, allen2025}. Importantly, entanglement need not arise solely from coherent unitary evolution: suitably chosen and adaptively processed measurements can also generate nonlocal correlations in many-body systems~\cite{bennett1993, popp2005, rajabpour2015, lin2023}. This measurement-induced entanglement (MIE) underlies measurement-based quantum computation~\cite{briegel2009}, enables rapid preparation of long-range entangled states~\cite{briegel2001, piroli2021, lee2022, zhu2023, cowsik2025}, and gives rise to novel non-equilibrium phases of matter~\cite{jian2021, chen2024, ippoliti2021, lavasani2021, klocke2025}. In particular, in hybrid unitary–measurement dynamics, competition between scrambling and projective measurements produces a measurement-induced phase transition (MIPT), 
across which the scaling of MIE changes from volume law to area law~\cite{li2019, skinner2019, szyniszewski2019, vasseur2019, nahum2020, zabalo2020, jian2020, bao2020, choi2020, turkeshi2020, alberton2021, sharma2022, block2022,sierant2022,fisher2023, poboiko2024, qian2024}.

Despite its conceptual and practical significance, directly characterizing MIE in experiments remains notoriously challenging~\cite{koh2023}. The central obstacle is post-selection: probing properties of a state conditioned on a specific measurement outcome requires repeating the experiment until that outcome reoccurs, an effort that grows exponentially with the number of measurements by Born's rule. Several scalable diagnostics have been proposed to circumvent this bottleneck, including purification of an entangled reference qubit~\cite{gullans2020a, gullans2020, noel2022, dehghani2023, hoke2023}, learnability of conserved quantities~\cite{agrawal2024} or of the pre-measurement state~\cite{ippoliti2024}, cross-entropy benchmarks~\cite{li2023, kamakari2025, qian2025} and other machine-learning proxies~\cite{turkeshi2022, qian2024b, kim2025}. Although these proxies successfully reflect the distinct behavior of MIE in different regimes and help reveal its critical behavior, they remain indirect witnesses rather than quantitative estimators of MIE itself. Recent works have made progress showing promise and limits~\cite{garratt2024, mcginley2024}. On the one hand, MIE can in principle be estimated without post-selection by leveraging quantum-classical correlations, but this requires prior knowledge of the underlying quantum dynamics and the accuracy of estimation hinges on the fidelity of classical simulations. On the other hand, Ref.~\cite{mcginley2024} proves that, without any knowledge of the pre-measurement state, no learning protocol can extract properties beyond ensemble averages using only a polynomial number of measurement shots. The practical limits of experimentally detectable MIE thus remain an open question.

Motivated by the recent success of data-driven quantum learning methods based solely on measurement records~\cite{huang2021, huang2022, du2025}, we ask in this Letter: under what conditions can MIE be learned with polynomial resources from data alone—i.e., without post-selection and without any prior knowledge of state preparation? To address this question, we study the entanglement generated between two distant qubits $A$ and $B$ after all other qubits in a many-body state are measured. Using measurement outcomes only, we train a transformer-based neural network in a self-supervised manner to estimate the post-measurement state on $AB \equiv A\cup B$ conditioned on a given measurement outcome. Instead of reconstructing MIE directly, we estimate the total entanglement entropy of $AB$, which we use as an uncertainty metric quantifying the learnability of MIE. We focus on random one-dimensional (1D) all-to-all circuits, in which we observe a clear learnability transition with increasing circuit depth. This transition is fundamentally a computational phase transition marked by a qualitative change in resource scaling: in the learnable phase, the uncertainty can be driven down with only polynomially many measurement shots and model parameters, whereas in the unlearnable phase it saturates or even grows, signaling that the required time and space complexity becomes exponential in the system size. We further probe this learnability transition in the presence of noise through simulations and experiments on IBM QPU $\text{ibm\_marrakesh}$, finding that the transition remains observable under realistic noise. Overall, our results demonstrate the power of data-driven methods to extract informative features of the post-measurement state that reveal MIE, while simultaneously clarifying the fundamental limitations of learning MIE from measurement data alone.
We note that a recent study~\cite{hou2025} investigated MIE learnability for GHZ and cluster states using a similar data-driven approach, where signatures of a learnability transition were observed for the latter upon tuning the measurement direction. 
In contrast, our work 
addresses generic states generated by random quantum circuits and provides an explicit characterization of the emergence of distinct learnable and unlearnable regimes.

\emph{Quantifying learnability---}We quantify the learnability of MIE as follows. Consider an $L$-qubit pure state $\ket{\psi}$. All qubits except two distant ones, $A$ and $B$, are measured in the computational basis, yielding outcome $m$. The corresponding post-measurement state on $AB$ is $\sigma_{AB,m}$, with reduced state $\sigma_{A,m}=\text{Tr}_{B}(\sigma_{AB,m})$. The MIE between $A$ and $B$ is defined as the average entanglement entropy $\mathbb{E}_{m}[S_{A,m}]$, $S_{A,m}=-\text{Tr}\left(\sigma_{A,m}\ln \sigma_{A,m}\right)$, where $\mathbb{E}_{m}$ denotes averaging over all measurement outcomes. Although evaluating $\mathbb{E}_{m}[S_{A,m}]$ exactly is difficult due to post-selection, two-sided bounds can be given once a classical estimator $\rho_{AB,m}$ of $\sigma_{AB,m}$ is provided, which is required to be a physical density matrix~\cite{garratt2024}. Nonnegativity of the relative entropy $D(\sigma_{A,m}||\rho_{A,m}) \equiv \text{Tr}[\sigma_{A,m}(\ln \sigma_{A,m}-\ln \rho_{A,m})] \geq 0$ implies the upper bound $S_{A,m} \leq S_{A,m}^{\text{QC}}$, where $S_{A,m}^{\text{QC}}\equiv-\text{Tr}\left(\sigma_{A,m}\ln\rho_{A,m}\right)$ is the quantum-classical entropy. Monotonicity under partial trace, $D(\sigma_{AB,m}||\rho_{AB,m}) \geq D(\sigma_{A,m}||\rho_{A,m})$, then yields:
\begin{equation}\label{bound}
S_{A,m}^{\text{QC}}\geq S_{A,m} \geq S_{A,m}^{\text{QC}} - S_{AB,m}^{\text{QC}},
\end{equation}
which utilizes $S_{AB,m} \geq 0$. Thus, the true entanglement $S_{A,m}$ lies within an interval of width $S_{AB,m}^{\text{QC}}$ for each outcome $m$. Nevertheless, directly computing $S_{AB,m}^{\text{QC}}$ requires complete knowledge of $\sigma_{AB,m}$. To access this quantity, we employ classical shadow tomography~\cite{huang2020}. Qubits $A$ and $B$ are each measured in a random single-qubit Pauli basis, specified by a rotation $U_{q}^{s}$ and computational-basis outcome $m_{q}$, giving the snapshot $\sigma_{AB,m}^{s}=\bigotimes_{q\in\{A,B\}}\big(3U_{q}^{s\dagger}\ket{m_{q}}\bra{m_{q}}U_{q}^{s}-I\big)$. These snapshots are unbiased estimators of $\sigma_{AB,m}$, with $\mathbb{E}_{s}[\sigma_{AB,m}^{s}]=\sigma_{AB,m}$, where $\mathbb{E}_{s}$ denotes averaging over the random bases~\footnote{Each snapshot $\sigma_{AB,m}^{s}$ is Hermitian with unit trace but generally not positive semidefinite, i.e., a formal classical-shadow estimator rather than a physical density matrix, unlike the estimator $\rho_{AB,m}$, which is physical by construction.}. We then define
\begin{equation}\label{delta}
\Delta\equiv\mathbb{E}_{m}\mathbb{E}_{s}\left[S_{AB,m}^{\text{SC}}\right]=\mathbb{E}_m\left[S_{AB,m}^{\text{QC}}\right],
\end{equation}
where $S_{AB,m}^{\text{SC}}\equiv -\text{Tr}(\sigma_{AB,m}^{s}\ln \rho_{AB,m})$ is the shadow-classical entropy. Random single-qubit Pauli rotations on $A$ and $B$, followed by computational-basis measurements of all qubits, jointly implement the averaging over both $m$ and classical snapshots, consolidating the two expectations and making $\Delta$ experimentally accessible. Since $S_{AB,m}^{\text{QC}}$ bounds the width of the interval containing the true entanglement, the quantity $\Delta$ directly tracks the uncertainty of MIE and therefore quantifies its learnability under a given learning scheme, i.e., a method mapping a state's measurement data to an estimator $\rho_{AB,m}$. It is also worth noting that $\Delta \geq \mathbb{E}_{m}[S_{AB,m}] \geq 0$, with the first inequality saturated when the estimator is exact, $\rho_{AB,m}=\sigma_{AB,m}$ for all $m$, and the second if every conditional post-measurement state $\sigma_{AB,m}$ is pure.

\emph{Machine-learning model}---Rather than estimating $\rho_{AB,m}$ by classically simulating the state-preparation dynamics, we learn it directly from the measurement data. We parameterize $\rho_{AB,m}$ with a neural network and train it to approximate the true post-measurement state $\sigma_{AB,m}$. To construct the loss function, we use the inequality $\mathbb{E}_{m}[\text{Tr}(\sigma_{AB,m}-\rho_{AB,m})^2] \geq 0$, which implies
\begin{equation}
 \begin{aligned}
 &\mathbb{E}_{m}\mathbb{E}_{s}\left[2\text{Tr}(\rho_{AB,m}\sigma_{AB,m}^{s}) - \text{Tr}(\rho_{AB,m}^2)\right]  
 \\
 &\leq \mathbb{E}_{m}\left[\text{Tr}\sigma_{AB,m}^2\right] \leq 1.
 \end{aligned}
\end{equation}
The first inequality is saturated when $\rho_{AB,m} = \sigma_{AB,m}$ for all $m$, and the second if $\sigma_{AB,m}$ is pure. This motivates the loss function:
\begin{equation}
\mathcal{L}(\theta) = -\mathbb{E}_{m}\mathbb{E}_{s}\left[2\text{Tr}(\rho_{AB,m}\sigma_{AB,m}^{s}) - \text{Tr}(\rho_{AB,m}^2)\right],
\end{equation}
where $\theta$ denotes the model parameters. We adopt this loss rather than directly targeting $\Delta$ because $\text{Tr}(\rho_{AB,m}\sigma_{AB,m}^{s})$ is bounded, while $S_{AB,m}^{\text{SC}}$ exhibits unbounded fluctuations~\cite{garratt2024}. 
Importantly, this approach uses only experimentally accessible data: the labels $\sigma_{AB,m}^{s}$ are directly reconstructed from measurements on $A$ and $B$, enabling fully self-supervised training.

\begin{figure}[t]
\begin{center}
\includegraphics[width=3.4in, clip=true]{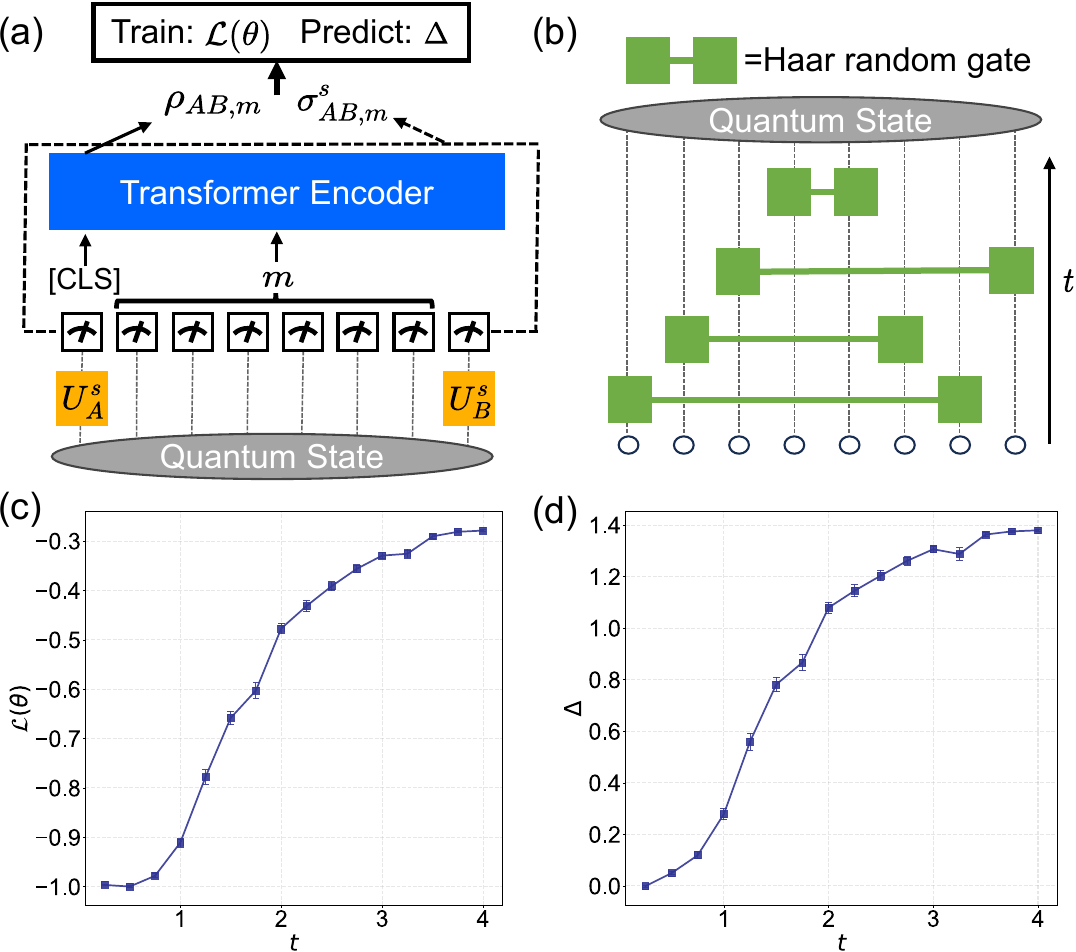}
\end{center}
\caption{Setup and signature of the learnability transition.  (a) Schematic workflow. Given a quantum state, random single-qubit rotations $U_{A}^{s}$ and $U_{B}^{s}$ are applied to qubits $A$ and $B$, after which all qubits are measured projectively in the computational basis. Measurement outcomes on all qubits except $A$ and $B$ are fed into a transformer encoder that outputs $\rho_{AB,m}$, while the outcomes on $A$ and $B$ are used to construct the classical snapshot $\sigma_{AB,m}^{s}$. The loss $\mathcal{L}(\theta)$ and uncertainty $\Delta$ are computed from these quantities. Notably, no prior knowledge of the state-preparation procedure is required. (b) Structure of the 1D all-to-all circuit. (c) $\mathcal{L}(\theta)$ as a function of circuit depth $t$ for $L=24$. Error bars denote the standard error over $M=50$ different circuit realizations. (d) $\Delta$ as a function of $t$ for $L=24$.}
\label{fig1}
\end{figure}

For the model architecture, we employ a transformer-based encoder consisting of multi-head self-attention layers and feed-forward blocks~\cite{vaswani2017, devlin2019}. Transformers efficiently capture long-range correlations, making them well-suited for modeling the nonlocal dependence of $\rho_{AB,m}$ on global measurement patterns. Each measurement outcome $m$, represented as a binary sequence of $\{+1, -1\}$, is augmented with a $[\text{CLS}]$ token at the beginning and encoded into high-dimensional representations. A specialized density matrix head then maps the $[\text{CLS}]$ token's hidden state to a $4 \times 4$ complex-valued density matrix, with hermiticity, positive semidefiniteness, and unit trace constraints automatically satisfied by construction. A schematic workflow is shown in Fig.~\ref{fig1}(a), with architectural details provided in~\cite{supp}. For a fixed quantum state, the training set contains $N_{m}$ measurement shots, which defines the quantum time complexity, while the model contains $N_{p}$ trainable parameters, reflecting the classical space complexity. Practical classical learnability without post-selection is feasible only if $N_{m}$ and $N_{p}$ do not grow exponentially with $L$. After training, we evaluate the learned estimator using an independent set of $N_{e}$ shots to compute $\Delta$ via Eq.~(\ref{delta}).

\emph{Learnability transitions}---With our learning scheme in place, we now examine the MIE learnability of concrete states. Specifically, we consider states generated by a 1D random all-to-all circuit, as illustrated in Fig.~\ref{fig1}(b). At each discrete time step $\delta t$, we apply $L\delta t$ Haar-random two-qubit gates between uniformly chosen pairs of qubits. For each depth $t$, we average over $M=50$ independently sampled circuit instances to capture average-case behavior. All training data are obtained from exact classical simulations of the circuit dynamics. We first train on a system of size $L=24$ using $N_{p}=7\times10^4$ model parameters and $N_{m} = 8\times10^4$ measurement shots. Additional hyperparameters are provided in~\cite{supp}. Fig.~\ref{fig1}(c) shows that the training loss $\mathcal{L}(\theta)$ converges to its lower bound $-1$ for shallow depths. As $t$ increases, the loss rises, indicating that the learned estimator $\rho_{AB,m}$ fails to reproduce the true post-measurement state. We then evaluate the uncertainty $\Delta$ on $N_{e} = 5\times10^3$ independent samples. As shown in Fig.~\ref{fig1}(d), $\Delta$ increases with $t$ and saturates near $2\ln(2)$, a first signature of two distinct regimes. This saturation suggests that at large $t$, the model outputs a maximally mixed state regardless of the measurement outcomes. Alternative loss functions may encourage pure-state predictions, but at large $t$ the predicted state becomes nearly orthogonal to the true one, still producing large $\Delta$~\cite{supp}.

\begin{figure}[t]
\begin{center}
\includegraphics[width=3.4in, clip=true]{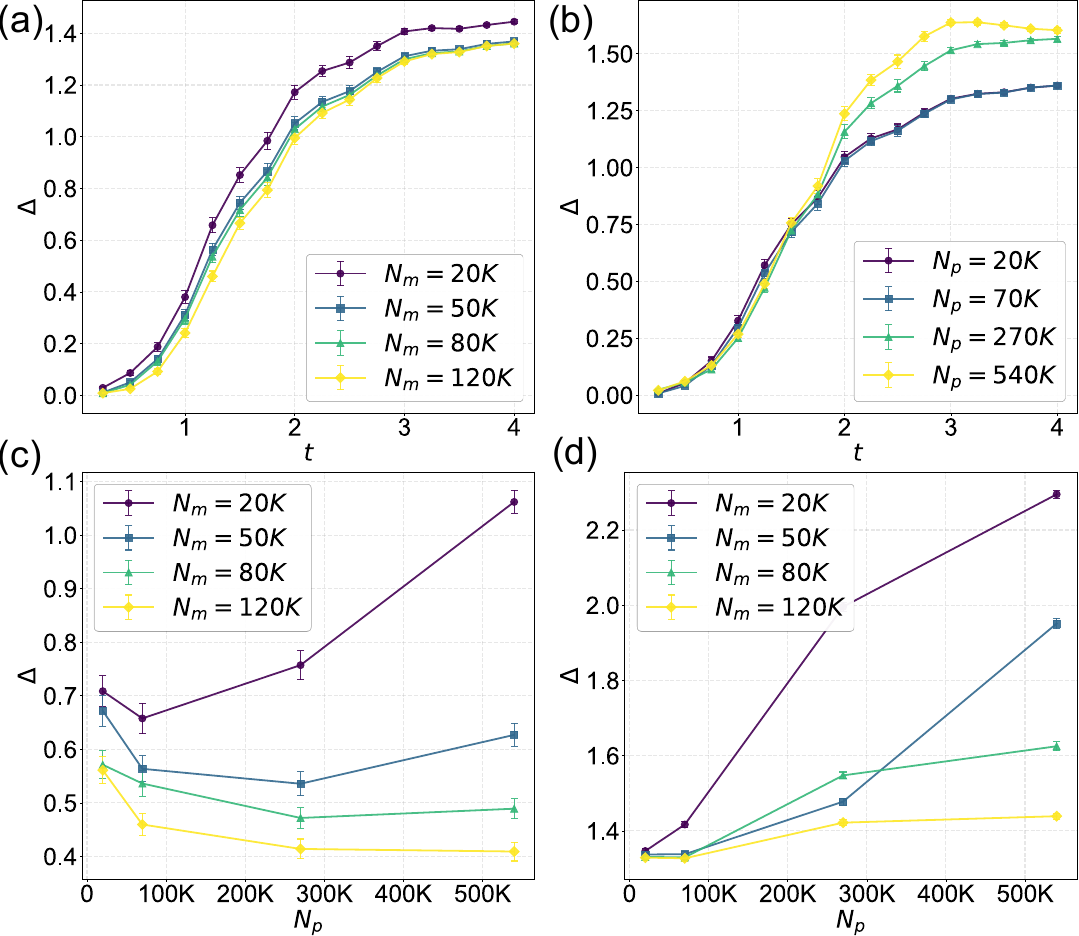}
\end{center}
\caption{Learnability transition in 1D all-to-all circuits for $L=20$. (a) $\Delta$ versus depth $t$ for varying $N_{m}$ with $N_{p} = 7\times10^4$ fixed. (b) $\Delta$ versus depth $t$ for varying $N_{p}$ with $N_{m}=5\times10^4$ fixed. (c,d) $\Delta$ as a function of $N_{m}$ and $N_{p}$ for representative depths $t=1.25$ and $t=3.5$, respectively.}
\label{fig2}
\end{figure}

To examine how the resources needed to learn the MIE scale, we first fix the system size at $L=20$ and vary $N_{p}$ and $N_{m}$. Detailed model architectures corresponding to different $N_{p}$ are provided in~\cite{supp}. Fixing $N_{p}=7\times10^4$ and varying $N_{m}$, Fig.~\ref{fig2}(a) shows that at small $t$, $\Delta$ decreases as data $N_{m}$ increases, indicating that additional samples improve the model's ability to extract MIE. In contrast, $\Delta$ saturates at large $t$ even as $N_{m}$ grows; the saturation seen in Fig.~\ref{fig1} is therefore intrinsic rather than a shortage of data, marking an unlearnable regime. Fixing $N_{m}=5\times10^4$ and varying $N_{p}$, Fig.~\ref{fig2}(b) shows that increasing $N_{p}$ improves performance at small $t$, while $\Delta$ saturates or even grows at large $t$. For excessively large $N_{p}$, $\Delta$ increases for all depths due to overfitting from insufficient training data. The combined dependence on $N_{p}$ and $N_{m}$ is shown in Fig.~\ref{fig2}(c) and Fig.~\ref{fig2}(d) for $\Delta (t=1.25)$ and $\Delta (t=3.5)$, respectively. At $t=1.25$, $\Delta$ clearly decreases with both resources (except in the strongly overparameterized regime where overfitting causes $\Delta$ to grow). By contrast, at $t=3.5$, $\Delta$ rapidly saturates and even grows despite the same increase in $N_{p}$ and $N_{m}$.

\begin{figure}[t]
\begin{center}
\includegraphics[width=3.4in, clip=true]{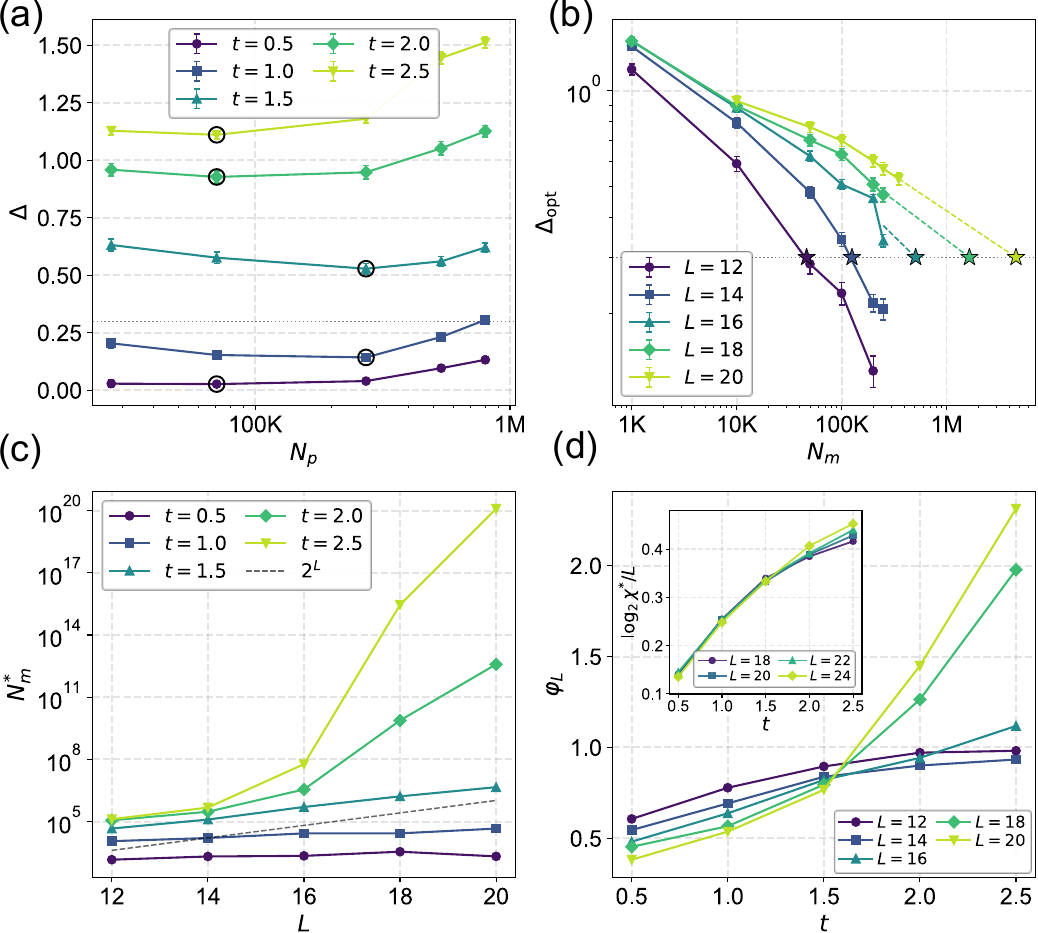}
\end{center}
\caption{Order parameter for the learnability transition in 1D all-to-all circuits. (a) $\Delta$ versus $N_{p}$ for different $t$ at $L=20$ with $N_{m}=3.5\times10^{5}$ fixed. The interior minimum (circle) defines $\Delta_{\mathrm{opt}}=\min_{N_{p}}\Delta$. (b) $\Delta_{\mathrm{opt}}$ versus $N_{m}$ for different $L$ at $t=1.5$. The smallest $N_{m}$ at which $\Delta_{\mathrm{opt}}$ falls below the threshold $\delta_{\mathrm{th}}=0.3$ (star) defines $N_{m}^{*}$. Results for other $\delta_{\mathrm{th}}$ are given in~\cite{supp}. (c) $N_{m}^{*}$ versus $L$ on a semilog scale, compared with the post-selection cost $2^{L}$. (d) $\varphi_{L}=\ln N_{m}^{*}/L$ versus $t$ for different $L$. The crossing locates the finite-depth transition. The inset shows the tensor-network order parameter $\log_2\chi^{*}/L$ versus $t$ for different $L$, where $\chi^{*}$ is the smallest bond dimension of an MPS compression of the exact final state for which $\Delta$ falls below $\delta_{\mathrm{th}}$.}
\label{fig3}
\end{figure}

To establish a genuine phase transition, we now vary the system size and construct an order parameter from the growth of the required resources. Since overfitting inflates $\Delta$ at large $N_{p}$, as seen in Fig.~\ref{fig2}, we first optimize over architectures: Fig.~\ref{fig3}(a) shows $\Delta$ as a function of $N_{p}$, whose interior minimum defines $\Delta_{\mathrm{opt}}=\min_{N_{p}}\Delta$ at each $(L,t,N_{m})$. From the dependence of $\Delta_{\mathrm{opt}}$ on $N_{m}$ shown in Fig.~\ref{fig3}(b), we extract the smallest measurement budget $N_m^{*}$ at which $\Delta_{\mathrm{opt}}$ drops below a fixed threshold, which sets the data required to learn the MIE at that depth. Fig.~\ref{fig3}(c) shows that the growth of $N_m^{*}$ with system size distinguishes the two phases: at small $t$, $N_m^{*}$ scales polynomially in $L$, while at large $t$ it grows exponentially in $L$. This motivates the order parameter $\varphi(t)=\lim_{L\to\infty}\ln N_m^{*}/L$, which vanishes in the learnable phase and is strictly positive in the unlearnable phase. As shown in Fig.~\ref{fig3}(d), the finite-size estimates $\varphi_{L}(t)=\ln N_m^{*}/L$ for different $L$ cross at a finite depth, locating the transition and reflecting the non-analyticity of $\varphi(t)$ in the thermodynamic limit.

We now compare this data-driven transition with the finite-depth teleportation transition of these random circuits~\cite{bao2024}, which marks a qualitative change in classical simulability~\cite{napp2022}. Below a critical depth, the dynamics is non-teleporting: measurement-induced correlations remain short-ranged, and the final state admits an efficient low-complexity classical representation. Above the critical depth, the circuit enters a teleporting regime, where long-range entanglement develops and the bond dimension required for an accurate matrix product state (MPS) representation grows exponentially with system size. To locate this boundary, we replace the neural-network estimator by an MPS-based one: the exact final state is compressed to bond dimension $\chi$ and used to compute $\rho_{AB,m}$ entering the same uncertainty $\Delta$. We define $\chi^{*}$ as the minimal bond dimension for which this MPS-based $\Delta$ falls below $\delta_{\mathrm{th}}$~\cite{supp}. As shown in the inset of Fig.~\ref{fig3}(d), the rescaled quantity $\log_2\chi^{*}/L$ exhibits a finite-size crossing at essentially the same depth as the learnability order parameter. Thus, within our numerical resolution, the onset of unlearnability coincides with the breakdown of an efficient classical representation of the underlying quantum state, despite the two transitions being independently defined. This agreement indicates that the learnability transition tracks the classical simulation boundary, providing a data-driven probe of quantum state complexity.

\begin{figure}[t]
\begin{center}
\includegraphics[width=3.4in, clip=true]{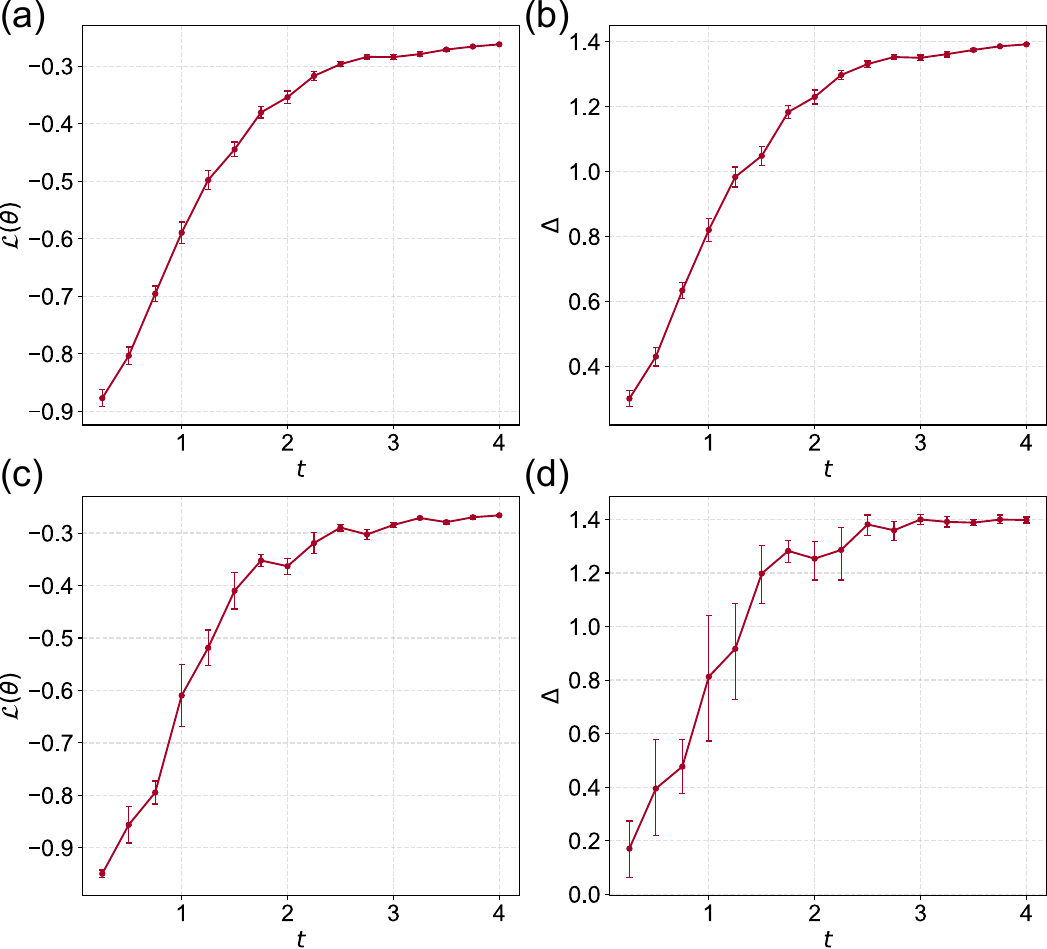}
\end{center}
\caption{Learnability transition in 1D all-to-all circuits with noise. (a,b) $\mathcal{L}(\theta)$ and $\Delta$ for $L=16$, obtained from noisy classical simulations using the Qiskit noise snapshot of the IBM QPU $\text{ibm\_brisbane}$. Here $N_{m}=2\times10^4$, $N_{p}=2\times10^4$, $N_{e}=5\times10^3$, and $M=50$. (c,d) $\mathcal{L}(\theta)$ and $\Delta$ for $L=20$, obtained from experiments on the IBM QPU $\text{ibm\_marrakesh}$ with sampling-compatible error mitigation (dynamical decoupling, Pauli and measurement twirling, and readout-calibrated shadows). Here $N_{m}=4\times10^4$, $N_{p}=7\times10^4$, $N_{e}=5\times10^3$, and $M=5$; the error bars in (d) are $95\%$ confidence intervals from a hierarchical bootstrap.}
\label{fig4}
\end{figure}

\emph{Noisy device---}The preceding analysis assumes ideal, noiseless circuit dynamics. However, in realistic experiments, noise introduces two key effects. First, each nominally identical state-preparation run experiences a different and unknown noise realization, making it harder for the model to extract consistent information from measurement records. Second, noise renders the post-measurement state $\sigma_{AB,m}$ mixed rather than pure, leading to $\mathbb{E}_{m}[S_{AB,m}]>0$. As discussed above, this finite entropy constitutes an intrinsic contribution to the uncertainty of MIE and cannot be eliminated by improving model accuracy. To examine these effects, we combine noisy numerical simulations with direct experiments on current superconducting quantum processors. As a first step, we adopt a noise model based on the Qiskit snapshot of the IBM QPU $\text{ibm\_brisbane}$ backend. For a 1D all-to-all circuit with $L=16$, Fig.~\ref{fig4}(a,b) show that the learnability transition remains visible, although $\Delta$ is systematically larger than that in the ideal case even at small depths $t$. This offset reflects the intrinsic uncertainty induced by noise. Due to the computational cost of noisy simulation at larger scales, we turn to direct experiments on the IBM QPU $\text{ibm\_marrakesh}$ for $L=20$~\cite{AbuGhanem2025}. Constrained by available experimental resources, we use $N_{m}=4\times10^4$, $N_{e}=5\times10^3$ and $M=5$, with all sampling-compatible error mitigation applied~\cite{supp}. The resulting data, Fig.~\ref{fig4}(c,d), show that $\Delta$ increases with $t$ and eventually saturates near $2\ln(2)$, demonstrating that the learnability transition persists on current noisy hardware. It is also noteworthy that $\Delta$ at small $t$ is lower than in the $\text{ibm\_brisbane}$ simulations, consistent with the lower noise level of $\text{ibm\_marrakesh}$ and the correspondingly reduced intrinsic uncertainty.

\emph{Discussions---}Our results have several implications for extracting MIE in practice and for understanding fundamental limits of quantum learning. In the shallow-depth regime, the network recovers the MIE from measurement data alone, with no prior knowledge of state preparation. This demonstrates that when the underlying state has low classical complexity, the structure relevant to MIE can be learned directly from data, suggesting the possibility of data-driven neural-network decoders rather than simulation-based decoders for quantum error correction~\cite{torlai2017, bausch2024, varbanov2025, hu2025, zhou2025}.
By contrast, at large depth the loss of learnability occurs where efficient classical representations break down. Although learning the MIE is not equivalent to simulating the full many-body wavefunction, this coincidence indicates that the same complexity growth underlying classical-simulation hardness also obstructs polynomial-resource data-driven recovery of MIE in our setting. This failure is physical rather than an artifact of the learning method, as we rule out finite model capacity, optimization difficulty, inductive bias, and the choice of loss function~\cite{supp}. This further suggests that in deep circuits within our setup, not only is substantial MIE generated, but crucially, this entanglement remains undetectable. Harnessing MIE in such regimes—rather than merely witnessing its presence—thus remains an important outstanding challenge.

Our work also suggests several interesting directions for future investigation. The present resource analysis accounts for quantum time complexity via $N_{m}$ and classical space complexity through $N_p$, but does not incorporate the classical time complexity of training itself. Including training overhead would sharpen the boundary of efficient MIE extraction and is essential for a complete complexity-theoretic characterization. In addition, our study has focused on states generated by random circuits, which represent an average-case scenario. 
Since physically motivated states—e.g., ground states of interacting Hamiltonians or states with specific symmetries—typically exhibit structure absent in random ensembles, exploring learnability transitions in such settings is a natural and important next step. We leave this to future work.

\begin{acknowledgments}
\emph{Acknowledgments}---We acknowledge Yi-Zhuang You and Pengfei Zhang for valuable discussions. This work is supported by the Natural Science Foundation of China through Grants No.~12350404 and No.~12174066, Quantum Science and Technology-National Science and Technology Major Project through Grant No.~2021ZD0302600, the Science and Technology Commission of Shanghai Municipality under Grants No.~23JC1400600 and No.~24LZ1400100, and is sponsored by the ``Shuguang Program'' supported by the Shanghai Education Development Foundation and Shanghai Municipal Education Commission.
\end{acknowledgments}

\end{document}

\title{Supplemental Material for ``Data-Driven Learnability Transition of Measurement-Induced Entanglement''}
\author{Dongheng Qian}
\affiliation{State Key Laboratory of Surface Physics and Department of Physics, Fudan University, Shanghai 200433, China}
\affiliation{Shanghai Research Center for Quantum Sciences, Shanghai 201315, China}
\author{Jing Wang}
\thanks{wjingphys@fudan.edu.cn}
\affiliation{State Key Laboratory of Surface Physics and Department of Physics, Fudan University, Shanghai 200433, China}
\affiliation{Shanghai Research Center for Quantum Sciences, Shanghai 201315, China}
\affiliation{Institute for Nanoelectronic Devices and Quantum Computing, Fudan University, Shanghai 200433, China}
\affiliation{Hefei National Laboratory, Hefei 230088, China}

%\date{\today}

\maketitle

\tableofcontents

\section{Definitions and estimator construction}
\label{sec:defs}

For clarity and completeness, we collect in this section the operational definitions of the objects introduced in the main text. We consider an $L$-qubit pure state $\ket{\psi}$; all of its qubits except two distant ones, $A$ and $B$, are measured in the computational basis, yielding an outcome $m$ and leaving $A$ and $B$ in the post-measurement state $\sigma_{AB,m}$, with reduced state $\sigma_{A,m}=\text{Tr}_{B}(\sigma_{AB,m})$ on $A$.

Our estimator of $\sigma_{AB,m}$, denoted $\rho_{AB,m}$, is obtained entirely by neural-network inference. The outcome $m$ is fed to a transformer encoder whose density-matrix head returns a $4\times4$ complex matrix; as detailed in Sec.~\ref{sec:arch}, this matrix is Hermitian, positive semidefinite, and of unit trace by construction, so that $\rho_{AB,m}$ is always a physical density matrix. The snapshots $\sigma_{AB,m}^{s}$, by contrast, are classical-shadow samples~\cite{huang2020}. Applying independent random single-qubit Pauli rotations $U_{A}^{s}$ and $U_{B}^{s}$ to $A$ and $B$ and measuring them in the computational basis yields
\begin{equation}
\sigma_{AB,m}^{s}=\bigotimes_{q\in\{A,B\}}\left(3\,U_{q}^{s\dagger}\ket{m_{q}}\bra{m_{q}}U_{q}^{s}-I\right),
\end{equation}
where $m_{q}$ is the outcome on qubit $q\in\{A,B\}$ and $I$ is the single-qubit identity. The snapshots are Hermitian, of unit trace, and unbiased, $\mathbb{E}_{s}[\sigma_{AB,m}^{s}]=\sigma_{AB,m}$, with $\mathbb{E}_{s}$ denoting the average over the random bases $s$. They are, however, generally not positive semidefinite and hence are not physical states.

From the physical estimator and these snapshots we define the quantum-classical and shadow-classical entropies
\begin{equation}
S_{AB,m}^{\text{QC}}\equiv-\text{Tr}\!\left(\sigma_{AB,m}\ln\rho_{AB,m}\right),\qquad S_{AB,m}^{\text{SC}}\equiv-\text{Tr}\!\left(\sigma_{AB,m}^{s}\ln\rho_{AB,m}\right),
\end{equation}
and the uncertainty metric as their average,
\begin{equation}
\Delta\equiv\mathbb{E}_{m}\mathbb{E}_{s}\!\left[S_{AB,m}^{\text{SC}}\right]=\mathbb{E}_{m}\!\left[S_{AB,m}^{\text{QC}}\right],
\end{equation}
where the second equality follows from $\mathbb{E}_{s}[\sigma_{AB,m}^{s}]=\sigma_{AB,m}$ and is what renders $\Delta$ experimentally accessible without post-selection. Because $S_{AB,m}^{\text{QC}}$ equals the width of the interval that brackets the true entanglement $S_{A,m}$ [Eq.~(1) of the main text], $\Delta$ measures the uncertainty of the measurement-induced entanglement (MIE) and thereby quantifies its learnability.

\section{Model architecture}
\label{sec:arch}
We employ an end-to-end learning framework based on a transformer encoder to map binary measurement sequences $\{+1, -1\}$ to $4\times4$ complex-valued density matrices that satisfy quantum mechanical constraints. The architecture consists of two main components: a BERT encoder that processes the input sequence and extracts the [CLS] token representation~\cite{devlin2019}, and a specialized density matrix generation head that produces the final output.

Our encoder follows the standard transformer architecture with a vocabulary size of 3~\cite{vaswani2017}, corresponding to the classification token and the two types of measurement outcomes. The input to the model is a binary measurement sequence, with a special [CLS] token prepended at the beginning. The input processing proceeds as follows: each token in the sequence is mapped through a trainable embedding layer to a $d_{h}$-dimensional vector, where $d_{h}$ denotes the hidden dimension of the model. Simultaneously, each position in the sequence is encoded by a learnable positional embedding, also of dimension $d_{h}$. Unlike the original transformer which uses fixed sinusoidal position encodings, we employ learnable positional embeddings that are jointly optimized during training, allowing the model to adapt its positional representations to the specific structure of quantum measurement sequences~\cite{devlin2019}. For the two-dimensional case, we use the same approach, as we expect the positional embeddings to capture the underlying two-dimensional structure of the measurement sequence. The token embeddings and positional embeddings are then element-wise added to form the input representation.

This input representation is processed by $N_{l}$ stacked transformer encoder layers. Each layer consists of two sub-components connected by residual connections: a multi-head self-attention mechanism and a position-wise feed-forward network.  We adopt the Pre-LN architecture where layer normalization is applied before each sub-component rather than after~\cite{radford2019}. In the multi-head self-attention mechanism, the input of dimension $d_{h}$ is linearly projected into queries, keys, and values for $N_h$ parallel attention heads, each operating on a subspace of dimension $d = d_{h} / N_h$. The scaled dot-product attention is computed independently for each head as $\text{Attention}(Q, K, V) = \text{softmax}(QK^T / \sqrt{d})V$. The outputs from all heads are concatenated and linearly projected back to dimension $d_{h}$, followed by dropout and residual connection with the input. Throughout the work, we choose $N_{h}=4$. The output of the attention sub-layer is then fed into the position-wise feed-forward network, which consists of two linear transformations with a ReLU activation in between. The first linear layer expands the representation from dimension $d_{h}$ to an intermediate dimension $d_{\text{f}}$, applies ReLU activation, and the second linear layer projects it back to dimension $d_{h}$. This is again followed by dropout and a residual connection. Dropout with a rate of $0.1$ is applied before each residual connection, and an independent attention dropout with rate $0.1$ is applied to the attention weights. After passing through all $N_{l}$ encoder layers, we extract the final hidden state corresponding to the [CLS] token position, which serves as a fixed-dimensional representation of the entire input sequence. This [CLS] representation, a vector of dimension $d_{h}$, is then fed into the density matrix generation head.

The density matrix head maps the [CLS] token representation from the BERT encoder to a complex-valued density matrix through an $AA^\dagger$ decomposition that automatically satisfies all quantum mechanical constraints. Initially, a linear projection maps the [CLS] representation to $32$ real-valued parameters, which are then reshaped into a complex matrix $A \in \mathbb{C}^{4 \times 4}$ by interpreting consecutive pairs of parameters as the real and imaginary components. The unnormalized density matrix is computed as $\rho_0 = AA^\dagger$, which is then trace-normalized to satisfy $\rho_{\text{norm}} = \rho_0 / \text{Tr}(\rho_0)$. Finally, we apply $\varepsilon$-mixing through a convex combination $\rho = (1 - \varepsilon)\rho_{\text{norm}} + \varepsilon I$, where $\varepsilon = 0.0001$ is a small mixing parameter and $I$ is the identity matrix. The $\varepsilon$-mixing step here prevents numerical instabilities associated with near-zero eigenvalues during training. This construction method ensures that all physical constraints required of quantum density matrices are rigorously satisfied. Hermiticity ($\rho = \rho^\dagger$) is automatically satisfied by the $AA^\dagger$ construction since $(AA^\dagger)^\dagger = (A^\dagger)^\dagger A^\dagger = AA^\dagger$. Positive semidefiniteness with all eigenvalues bounded below by $\varepsilon$ is explicitly ensured through the $\varepsilon$-mixing step, which expresses the output as a convex combination of the learned density matrix and the maximally mixed state. The unit trace constraint $\text{Tr}(\rho) = 1$ is enforced through explicit trace normalization.

The majority of trainable parameters reside in the transformer encoder. To construct models with varying parameter numbers, we adjust three key architectural hyperparameters: the hidden dimension $d_{h}$, the feed-forward intermediate dimension $d_{\text{f}}$, and the number of encoder layers $N_{l}$. We constructed fifteen models spanning $\sim10^4$ to $\sim10^7$ parameters, and the relationship between model size and architectural configuration is summarized in Table~\ref{table1}.

\begin{table}[h]
\centering
\caption{Model configurations for different parameter numbers. The scan spans fifteen architectures from $\sim10^{4}$ to $\sim10^{7}$ parameters; the listed $N_{p}$ is a representative value, as the exact count varies by $\lesssim2\%$ with system size through the positional embeddings.} \label{table1}
\renewcommand{\arraystretch}{1.2}
\begin{tabular*}{4.5in}
{@{\extracolsep{\fill}}cccc}
\hline
Model Parameters $N_p$ & $d_{h}$ & $d_{\text{f}}$ & $N_l$ \\
\hline
10K & 32 & 64 & 1 \\
20K & 32 & 64 & 2 \\
27K & 32 & 128 & 2 \\
70K & 64 & 128 & 2 \\
100K & 64 & 256 & 2 \\
140K & 64 & 128 & 4 \\
200K & 64 & 256 & 4 \\
270K & 128 & 256 & 2 \\
400K & 128 & 512 & 2 \\
540K & 128 & 256 & 4 \\
800K & 128 & 512 & 4 \\
1.2M & 128 & 512 & 6 \\
2.1M & 256 & 512 & 4 \\
3.2M & 256 & 512 & 6 \\
12.6M & 512 & 1024 & 6 \\
\hline
\end{tabular*}
\end{table}

\section{Training details}

We optimize the model using the AdamW optimizer with an initial learning rate of $\eta_0 = 5 \times 10^{-4}$ and weight decay of $0.01$, which implements $L_2$-regularization through a decoupled weight decay parameter separate from the gradient-based updates. The learning rate follows a cosine annealing schedule with warmup: during an initial warmup phase comprising 10\% of total training steps, the learning rate increases linearly from zero to $\eta_0$, after which it gradually decreases following a cosine curve to a minimum value of $\eta_{\text{min}} = 10^{-5}$ to improve convergence to sharp minima. We apply global gradient norm clipping with a threshold of 1.0 to prevent gradient explosion, particularly important when training with complex-valued operations. Weight decay regularization is applied to all parameters except biases and layer normalization parameters, following standard practice in transformer training. The entire model is implemented in pure PyTorch, with no reliance on pre-trained weights; all components are trained from scratch. All the training is conducted on one 24GB NVIDIA 4090 GPU, one 48GB Ada6000 GPU and one 80GB A100 GPU. For all the cases, the batch size is set to match the total number of data. Throughout the study, we set the number of epochs to 500. 

\section{Asymptotic resource scaling}
\label{sec:scaling}

\subsection{Learnability order parameter}

\begin{figure}[tbp]
\begin{center}
\includegraphics[width=\textwidth, clip=true]{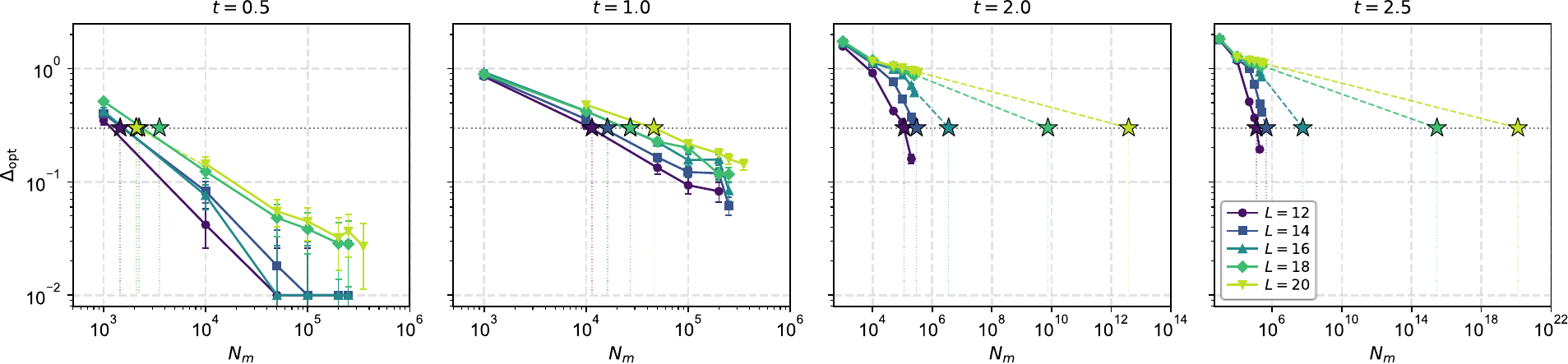}
\end{center}
\caption{Transparent extraction of the minimum data budget $N_m^{*}$. $\Delta_{\mathrm{opt}}=\min_{N_p}\Delta$ as a function of $N_m$ for $L=12$--$20$, at depths (a) $t=0.5$, (b) $t=1.0$, (c) $t=2.0$, and (d) $t=2.5$; the representative depth $t=1.5$ appears in Fig.~3 of the main text. The dotted line marks the threshold $\delta_{\mathrm{th}}=0.3$ and the stars mark $N_m^{*}$, obtained by interpolation when the threshold is bracketed and by power-law extrapolation (dashed) otherwise.}
\label{supp_fig1}
\end{figure}

We characterize the learnability transition quantitatively through the minimum data budget $N_m^{*}$ required to learn the MIE. For each system size $L$ and depth $t$ we first optimize over the model architecture, defining $\Delta_{\mathrm{opt}}(N_m)=\min_{N_p}\Delta$ at fixed measurement budget; the minimization uses different architectures per point with $N_p\in[1\times10^4,\,1.26\times10^7]$, ensuring that $\Delta_{\mathrm{opt}}$ is not limited by model capacity (Sec.~\ref{sec:algo}). We then define $N_m^{*}(L,t)$ as the smallest $N_m$ for which $\Delta_{\mathrm{opt}}$ drops below the threshold $\delta_{\mathrm{th}}=0.3$. Because $\Delta_{\mathrm{opt}}$ is sampled on a discrete grid of budgets, $N_m^{*}$ is obtained by one of three procedures, made transparent in Fig.~\ref{supp_fig1}: linear interpolation of $\Delta_{\mathrm{opt}}$ against $\ln N_m$ when the threshold is bracketed by adjacent budgets; downward extrapolation of a power-law fit ($\ln\Delta_{\mathrm{opt}}$ linear in $\ln N_m$) through the four smallest budgets when even the smallest already lies below $\delta_{\mathrm{th}}$; and upward extrapolation of the same fit through the four largest budgets when the threshold is not reached at any accessible budget. This last case---an $N_m^{*}$ extrapolated beyond every budget we can afford---is itself the operational signature of unlearnability.

We next ask how $N_m^{*}$ grows with system size. At small depth the growth is polynomial: fitting $N_m^{*}\sim L^{\alpha}$ over $L\in\{12,\dots,20\}$ returns a small and stable exponent, $\alpha\approx1.1$. At large depth this form breaks down---the apparent exponent grows steeply with depth, reaching $\alpha\approx70$ by $t=2.5$---and $N_m^{*}$ grows exponentially in $L$ instead. To capture this change in a single, size-independent quantity, we define the order parameter
\begin{equation}
\varphi(t)=\lim_{L\to\infty}\frac{\ln N_m^{*}(L,t)}{L},
\end{equation}
which vanishes when $N_m^{*}$ grows polynomially and is positive when it grows exponentially. At finite size we track $\varphi_L(t)=\ln N_m^{*}(L,t)/L$; as shown in Fig.~3(d) of the main text, $\varphi_L$ decreases with $L$ at small depth and increases with $L$ at large depth, so the curves for different $L$ cross at the depth separating the two phases. Averaging this crossing over all size pairs with both $L\ge14$ gives $t_c=1.57\pm0.04$, tightening to $1.55\pm0.01$ for pairs with both $L\ge16$; the quoted spread is the standard deviation of the crossing over pairs, reflecting the residual finite-size drift rather than a fit uncertainty.

The location is robust to the one free parameter in its definition, the threshold $\delta_{\mathrm{th}}$: varying $\delta_{\mathrm{th}}$ by a factor of two, from $0.2$ to $0.4$, shifts $t_c$ only within $1.5$--$1.6$, with the drift shrinking as the smallest included size grows (Table~\ref{tab:dthrobust}). This $t_c\approx1.5$ agrees with the finite-depth teleportation transition~\cite{bao2024}. Pinning $t_c$ down further by a data collapse is, however, limited by the resolution in depth: two-qubit gates are applied $L\,\delta t$ at a time step $\delta t$, so a depth $t$ corresponds to $Lt$ gates, and evaluating $\varphi_L$ at a shared depth across sizes requires $Lt$ to be an integer for every $L$ at once. At the small, classically simulable sizes $L\in\{14,16,18,20\}$ this forces a coarse step $\delta t=1/2$, leaving only a few depths in the critical window around $t_c$.

\begin{table}[h]\centering
\caption{Robustness of the crossing depth $t_c$ to the threshold $\delta_{\mathrm{th}}$ that defines $N_m^{*}$. Here $t_c$ is the mean pairwise $\varphi_L$ crossing over system-size pairs with both $L\ge14$ (resp.\ $L\ge16$), and the uncertainty is the standard deviation over pairs. Across a factor of two in $\delta_{\mathrm{th}}$, $t_c$ stays within $1.5$--$1.6$, and the residual drift shrinks as the smallest included $L$ grows.}\label{tab:dthrobust}
\renewcommand{\arraystretch}{1.2}
\begin{tabular*}{3.0in}
{@{\extracolsep{\fill}}ccc}
\hline
$\delta_{\mathrm{th}}$ & $t_c$ ($L\ge14$) & $t_c$ ($L\ge16$) \\
\hline
0.20 & $1.51\pm0.04$ & $1.53\pm0.01$ \\
0.25 & $1.53\pm0.01$ & $1.54\pm0.01$ \\
0.30 & $1.57\pm0.04$ & $1.55\pm0.01$ \\
0.35 & $1.60\pm0.09$ & $1.55\pm0.02$ \\
0.40 & $1.63\pm0.09$ & $1.58\pm0.01$ \\
\hline\hline
\end{tabular*}
\end{table}

\begin{figure}[tbp]
\begin{center}
\includegraphics[width=0.45\textwidth, clip=true]{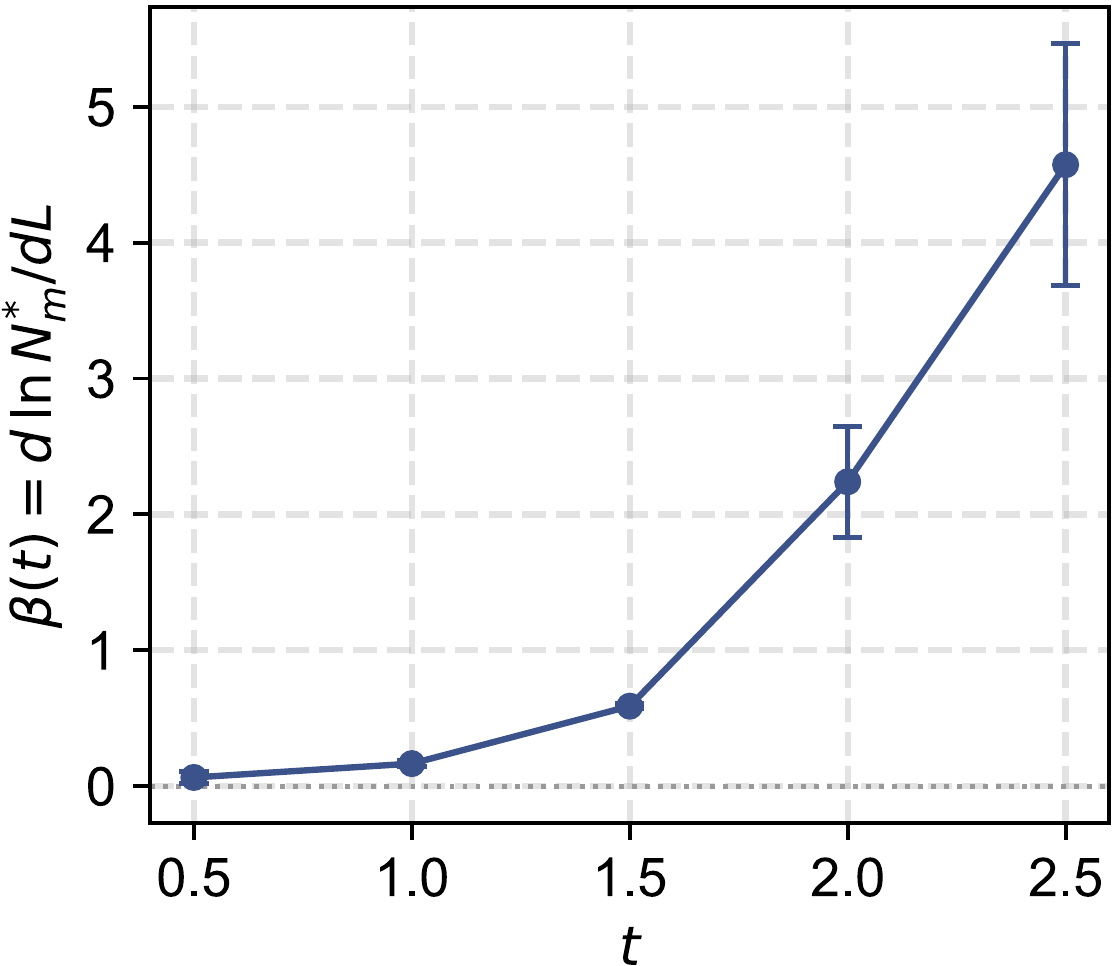}
\end{center}
\caption{Exponential growth rate of the data budget $N_m^{*}$. The rate $\beta(t)\equiv d\ln N_m^{*}/dL$, obtained as the slope of a linear fit of $\ln N_m^{*}$ against $L$ over $L\in\{12,\dots,20\}$, versus depth $t$; error bars give the standard error of the fitted slope. In the learnable phase $N_m^{*}$ grows only polynomially in $L$ and $\beta$ is small; through the transition $\beta$ rises monotonically to the large positive values of the unlearnable phase, where $N_m^{*}\sim e^{\beta L}$ grows exponentially with system size.}
\label{supp_fig2}
\end{figure}

Finally, we examine the transition from a complementary perspective, through the exponential growth rate $\beta(t)$---the slope of a linear fit of $\ln N_m^{*}$ against $L$ over $L\in\{12,\dots,20\}$---which measures how fast the data budget grows with system size and provides additional evidence for the transition. As shown in Fig.~\ref{supp_fig2}, $\beta(t)$ is small in the learnable phase, where $N_m^{*}$ grows only polynomially, and rises monotonically through the transition to the large positive values of the unlearnable phase, where the growth is exponential, tracing how the order parameter grows with depth. Taken together, these diagnostics---the order parameter $\varphi(t)$ and its finite-size crossing, the robustness of $t_c$ to the threshold, and the growth rate $\beta(t)$---consistently indicate that the learnability transition is present near $t_c\approx1.5$, although an accurate determination of its location requires more extensive simulations at larger sizes.

\subsection{Classical-simulation order parameter}

We now ask whether the state itself undergoes a transition in how hard it is to simulate classically, and whether that transition sits at the same depth as the learnability transition. We probe this by replacing the learned estimator with a tensor-network one, a matrix-product state (MPS) of bounded bond dimension $\chi$, constructed in two ways. In the first, we evolve the circuit exactly to the final state and truncate it to the best bond-$\chi$ MPS by a sweep of singular-value decompositions; at $\chi=2^{L/2}$ the MPS is exact and $\Delta$ vanishes at every depth, so lowering $\chi$ measures how much bond dimension the MIE actually requires. In the second, we evolve an MPS from the start by time-evolving block decimation (TEBD), applying the gates one at a time---the long-range ones through swap networks---and truncating back to bond $\chi$ after each, so the truncation error accumulates as in a genuine classical simulation. Both constructions rely on knowledge that the data-driven protocol never has: the first is handed the exact final state, the second the exact circuit, whereas the network sees only measurement outcomes.

\begin{figure}[tbp]
\begin{center}
\includegraphics[width=\textwidth, clip=true]{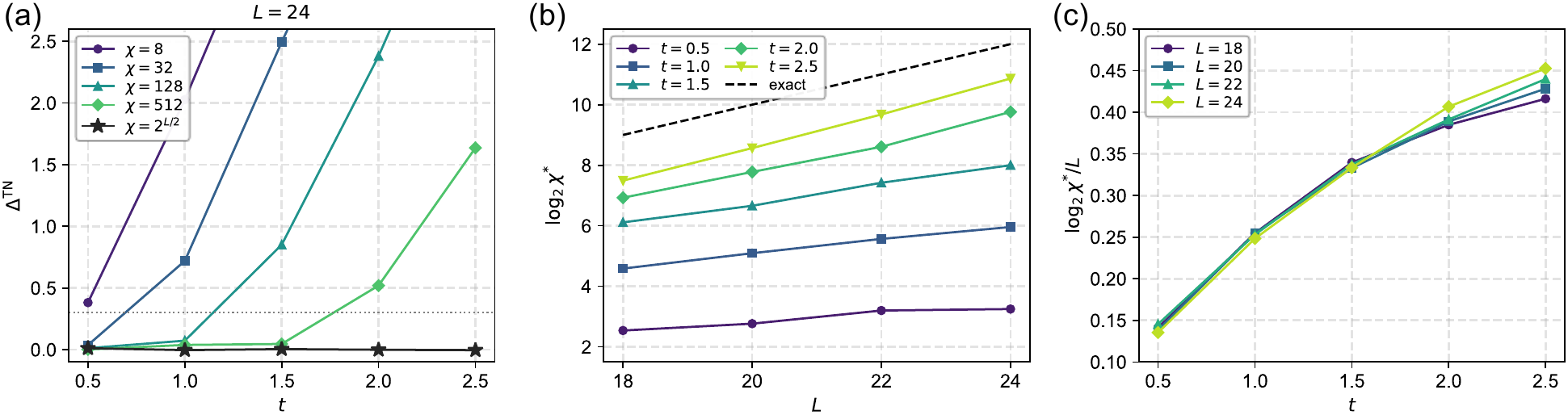}
\end{center}
\caption{Tensor-network estimator obtained by optimal compression of the exactly evolved final state. (a) $\Delta^{\mathrm{TN}}$ versus depth $t$ for several bond dimensions $\chi$ at $L=24$; at $\chi=2^{L/2}$ the uncertainty vanishes at all depths. (b) $\log_2\chi^{*}$ versus $L$ for several depths, approaching the exact ceiling $L/2$ above the transition. (c) the rescaled bond dimension $\log_2\chi^{*}/L$ versus $t$ for several $L$, crossing near $t_c\approx1.5$.}
\label{supp_fig3}
\end{figure}

\begin{figure}[tbp]
\begin{center}
\includegraphics[width=\textwidth, clip=true]{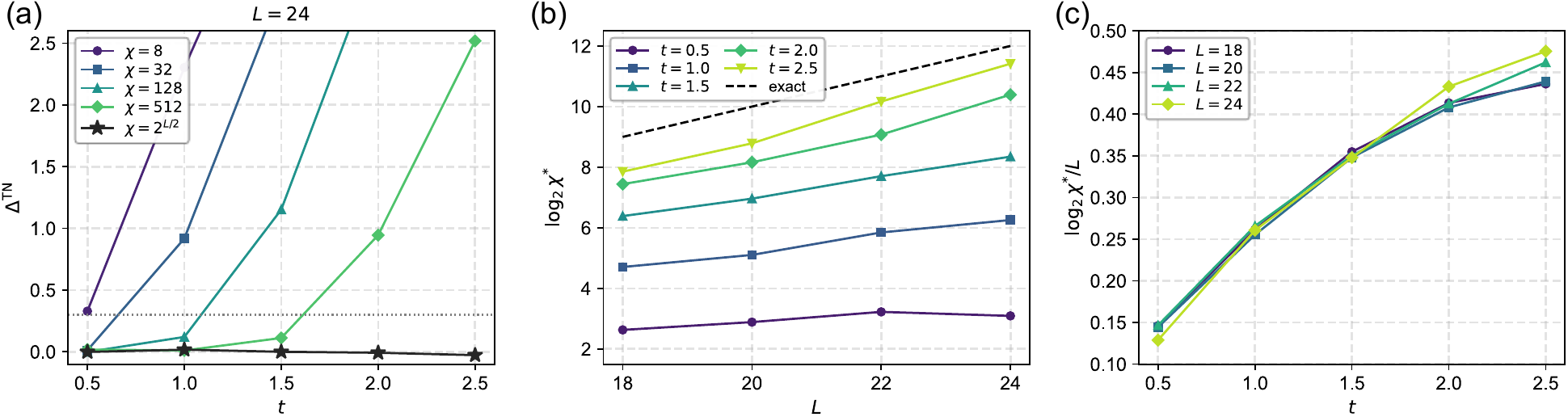}
\end{center}
\caption{Tensor-network estimator obtained by gate-by-gate TEBD with swap networks, given the exact circuit. Panels as in Fig.~\ref{supp_fig3}: (a) $\Delta^{\mathrm{TN}}$ versus $t$ for several $\chi$ at $L=24$; (b) $\log_2\chi^{*}$ versus $L$, approaching $L/2$ above the transition; (c) $\log_2\chi^{*}/L$ versus $t$, crossing near $t_c\approx1.5$.}
\label{supp_fig4}
\end{figure}

For each construction we extract, over $L=18$--$24$, the minimal bond dimension $\chi^{*}(L,t)$ that brings $\Delta$ below $\delta_{\mathrm{th}}$. Because the bond dimension is fixed by the entanglement across the cut, $\chi^{*}$ measures how hard the state is to represent, and its rescaled form $\log_2\chi^{*}/L$ serves as a classical-simulation order parameter analogous to $\varphi_L$. As shown in Figs.~\ref{supp_fig3} and~\ref{supp_fig4}, $\chi^{*}$ grows only weakly with $L$ (effectively polynomial) while the state remains simulable and approaches the maximal $2^{L/2}$ once it does not, and the rescaled curves for different $L$ cross near $t\approx1.5$. A classical-simulation-hardness transition therefore does occur, and its crossing lies very close to that of the learnability transition. That the MIE becomes unlearnable near where the state becomes hard to simulate classically is consistent with the two being the same underlying computational phase transition.

\section{Algorithmic versus physical limits}
\label{sec:algo}

The loss of learnability at large depth could in principle reflect a limitation of the learning algorithm rather than a physical one. We examine the standard algorithmic confounders---finite model capacity, optimization difficulty, inadequate inductive bias, and the choice of loss function---and find the transition robust to all of them.

\subsection{Model capacity}

\begin{figure}[tbp]
\begin{center}
\includegraphics[width=\textwidth, clip=true]{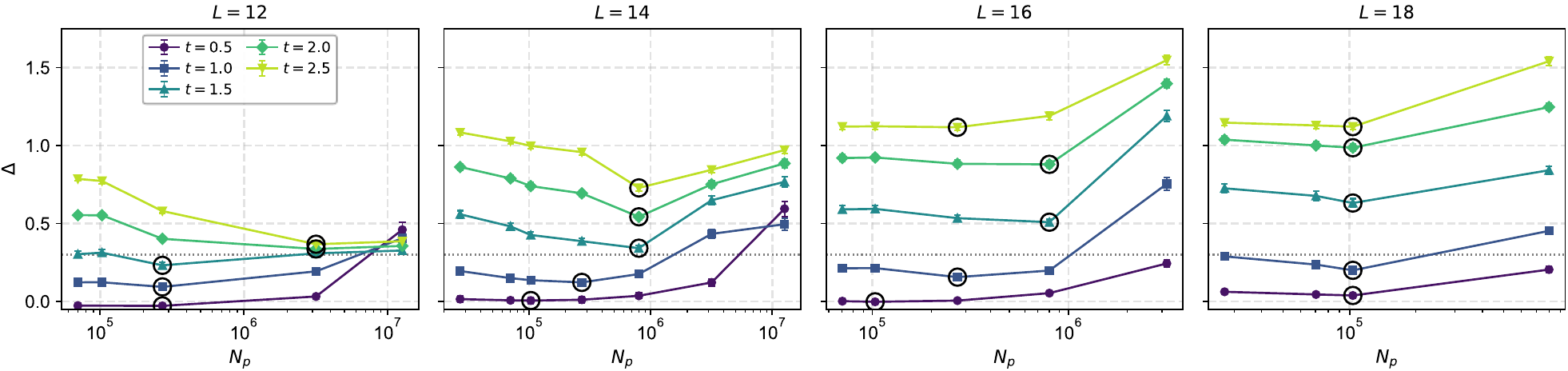}
\end{center}
\caption{$\Delta$ as a function of $N_{p}$ at a fixed data budget $N_{m}=10^{5}$, for several depths $t$, at (a) $L=12$, (b) $L=14$, (c) $L=16$, and (d) $L=18$. The optimum (open circle) never lies at the largest $N_{p}$---$\Delta$ rises again there through overfitting---so increasing the number of parameters does not reduce $\Delta$.}
\label{supp_fig5}
\end{figure}

A first concern is that the network is simply too small at large depth. As shown in Fig.~\ref{supp_fig5}, however, $\Delta$ rises again at large $N_p$ through overfitting, so its minimum never lies at the largest model. Increasing capacity therefore does not reduce $\Delta$: capacity is not the bottleneck.

\subsection{Optimization}

\begin{figure}[htbp]
\begin{center}
\includegraphics[width=0.45\textwidth, clip=true]{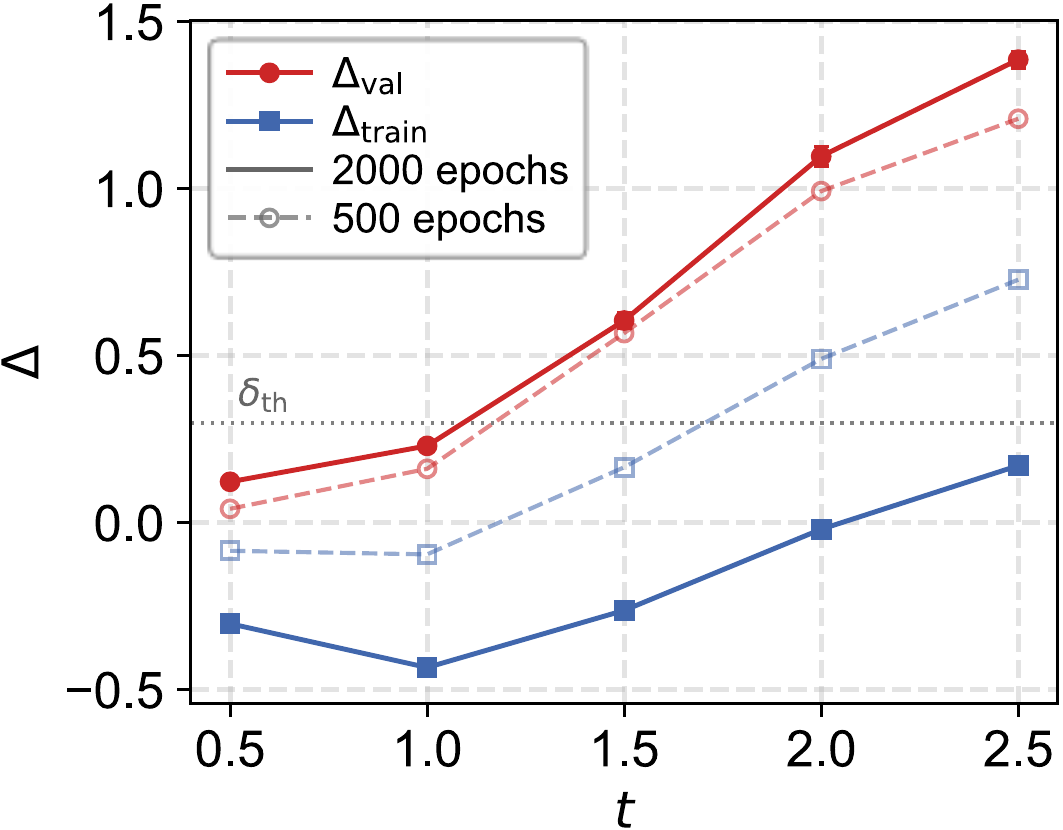}
\end{center}
\caption{The uncertainty $\Delta$ evaluated on the training shots, $\Delta_{\mathrm{train}}$, and on held-out shots, $\Delta_{\mathrm{val}}$, as a function of depth $t$ at $L=20$ and $N_{m}=2.5\times10^5$, with training extended to $2000$ epochs. The optimizer drives $\Delta_{\mathrm{train}}\to0$ even in the unlearnable phase, while $\Delta_{\mathrm{val}}$ is essentially unchanged---a pure generalization gap.}
\label{supp_fig6}
\end{figure}

The optimizer might instead be failing to reach the loss minimum. To rule this out we evaluate the uncertainty $\Delta$ on the training shots, $\Delta_{\mathrm{train}}$, and on held-out shots, $\Delta_{\mathrm{val}}$---the latter being the $\Delta$ reported throughout---while training well beyond our standard budget, to $2000$ epochs at $L=20$ and $N_m=2.5\times10^5$. As shown in Fig.~\ref{supp_fig6}, even in the unlearnable phase the optimizer drives $\Delta_{\mathrm{train}}$ nearly to zero---so it is not stuck at a poor minimum---while $\Delta_{\mathrm{val}}$ does not decrease, even rising slightly. The failure is therefore a generalization gap, not an optimization one: the network fits the shots it is trained on but cannot generalize to unseen ones, which is precisely the unlearnability of the MIE.

\subsection{Inductive bias}

\begin{figure}[tbp]
\begin{center}
\includegraphics[width=\textwidth, clip=true]{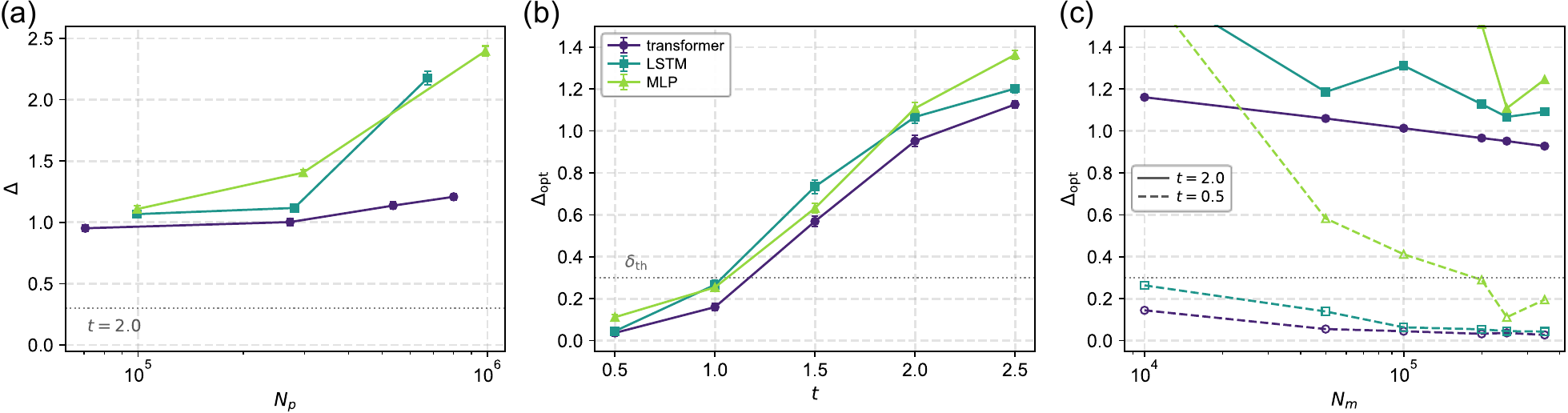}
\end{center}
\caption{The transition is invariant to the inductive bias, shown for a transformer, a bidirectional LSTM, and a flat multilayer perceptron at $L=20$. (a) $\Delta$ versus $N_p$ past the transition ($t=2.0$, $N_m=2.5\times10^5$); $\Delta$ stays above $\delta_{\mathrm{th}}$ and rises at large $N_p$ through overfitting for every architecture. (b) $\Delta_{\mathrm{opt}}=\min_{N_p}\Delta$ versus depth $t$ at $N_m=2.5\times10^5$; all three rise through $\delta_{\mathrm{th}}$ at the same depth. (c) $\Delta_{\mathrm{opt}}$ versus the data budget $N_m$ at a learnable depth ($t=0.5$, dashed) and past the transition ($t=2.0$, solid); more data lowers $\Delta_{\mathrm{opt}}$ below $\delta_{\mathrm{th}}$ at $t=0.5$ but not at $t=2.0$. The transformer attains the lowest $\Delta$ throughout.}
\label{supp_fig7}
\end{figure}

The result might also hinge on the transformer's attention-based prior. To test this, we keep the density-matrix head, the loss, and the optimizer fixed and replace the transformer backbone by two architectures with qualitatively different inductive biases. The first is a bidirectional long short-term memory (LSTM) network~\cite{hochreiter1997,schuster1997}: the embedded record is processed recurrently in both directions and the hidden states are mean-pooled over the sequence before the head, so correlations are accumulated sequentially rather than through global self-attention. The second is a flat multilayer perceptron that concatenates the token embeddings at all positions into a single vector and passes it through several fully connected ReLU layers---a backbone with neither recurrence nor attention that treats the measurement record as an unstructured feature vector. We compare all three at $L=20$ along the three resource axes of Fig.~\ref{supp_fig7}. Fig.~\ref{supp_fig7}(a) shows $\Delta$ versus $N_p$ past the transition ($t=2.0$, $N_m=2.5\times10^5$): for every architecture $\Delta$ stays well above $\delta_{\mathrm{th}}$ and rises again at large $N_p$ through overfitting, so more parameters do not help. Fig.~\ref{supp_fig7}(b) shows $\Delta_{\mathrm{opt}}=\min_{N_p}\Delta$ versus depth $t$ at $N_m=2.5\times10^5$: all three rise through $\delta_{\mathrm{th}}$ at essentially the same depth, marking the same transition. Fig.~\ref{supp_fig7}(c) shows $\Delta_{\mathrm{opt}}$ versus $N_m$ at a learnable depth ($t=0.5$, dashed) and past the transition ($t=2.0$, solid): at $t=0.5$ more data drives $\Delta_{\mathrm{opt}}$ below $\delta_{\mathrm{th}}$ for all three, while at $t=2.0$ it plateaus far above $\delta_{\mathrm{th}}$, so more data does not help either. The transformer attains the lowest $\Delta$ throughout---which is why we adopt it in the main text---yet even it fails to recover the MIE past the transition. The transition is therefore invariant under three qualitatively distinct inductive biases, and is not an artifact of the transformer's attention-based prior.

\subsection{Loss function}

\begin{figure}[htbp]
\begin{center}
\includegraphics[width=0.7\textwidth, clip=true]{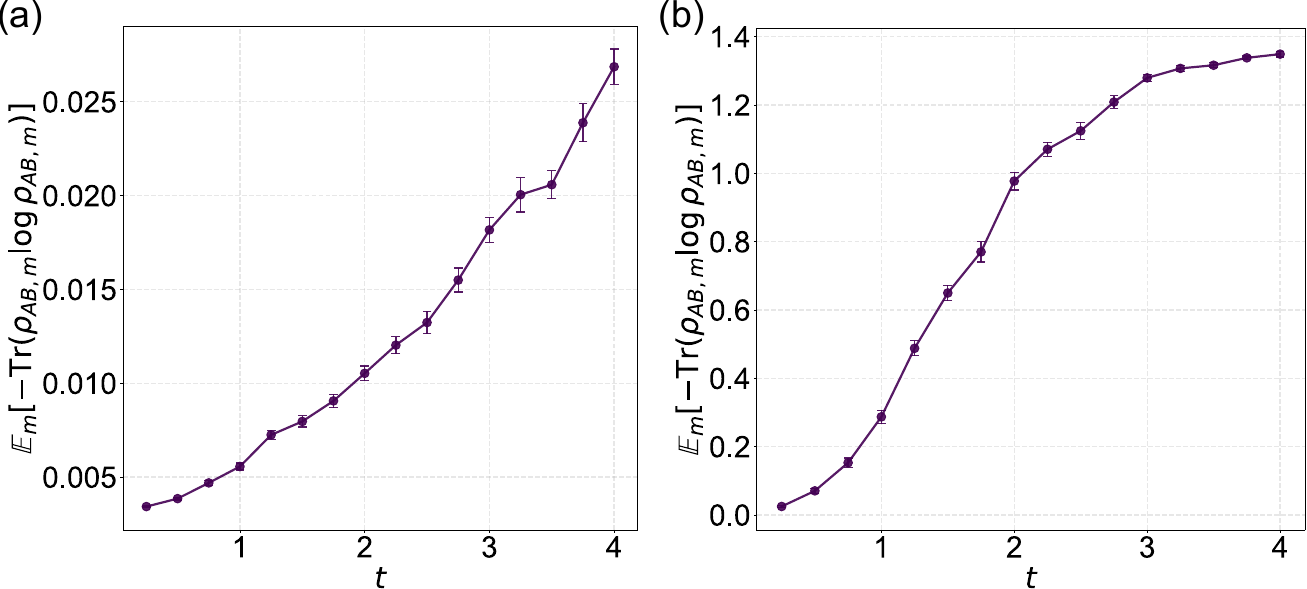}
\end{center}
\caption{Comparison of loss functions for random one-dimensional all-to-all circuits with $L=20$, $N_{m}=8\times10^4$, $N_{p}=7 \times 10^4$, and $N_{e}=5\times 10^3$. Error bars represent the standard error over $M=50$ circuit realizations. (a) Model trained with $\mathcal{L}_{1}(\theta)$. (b) Model trained with $\mathcal{L}(\theta)$. }
\label{supp_fig8}
\end{figure}

Finally, the transition might be specific to our loss $\mathcal{L}(\theta)$. Alternative losses are possible. For example, consider the inequality:
\begin{equation}
\mathbb{E}_{m}\mathbb{E}_{s}\left[\text{Tr}(\rho_{AB,m}\sigma_{AB,m}^{s})\right] = \mathbb{E}_{m}\left[\text{Tr}(\rho_{AB,m}\sigma_{AB,m})\right] \leq 1,
\end{equation}
where equality holds if and only if $\rho_{AB,m} = \sigma_{AB,m}$ for every $m$, assuming $\sigma_{AB,m}$ is pure. This motivates the following alternative loss function:
\begin{equation}
\mathcal{L}_{1}(\theta) = - \mathbb{E}_{m}\mathbb{E}_{s}[\text{Tr}(\rho_{AB,m}\sigma_{AB,m}^{s})].
\end{equation}
This loss applies only when the pre-measurement state is pure.  It differs from $\mathcal{L}(\theta)$ in two respects. First, it requires fewer samples to estimate accurately: its variance, dominated by $\text{Tr}(\rho_{AB,m}\sigma_{AB,m}^s)$, is about four times smaller than that of $\mathcal{L}(\theta)$.  Second,  $\mathcal{L}_{1}(\theta)$ favors pure-state outputs. Since $\text{Tr}(\rho_{AB,m}\sigma_{AB,m}^{s})$ is linear in $\rho_{AB,m}$, its maximum over the convex set of density matrices is attained at a boundary point---a pure state, the leading eigenvector of $\sigma_{AB,m}^{s}$---whereas the quadratic term in $\mathcal{L}(\theta)$ penalizes purity and so admits mixed states. As shown in Fig.~\ref{supp_fig8}, the entanglement entropy of $\rho_{AB,m}$ stays small under $\mathcal{L}_{1}(\theta)$ at all $t$, while under $\mathcal{L}(\theta)$ it becomes near-maximal at large $t$, as $\rho_{AB,m}$ approaches the maximally mixed state.

\begin{figure}[tbp]
\begin{center}
\includegraphics[width=0.7\textwidth, clip=true]{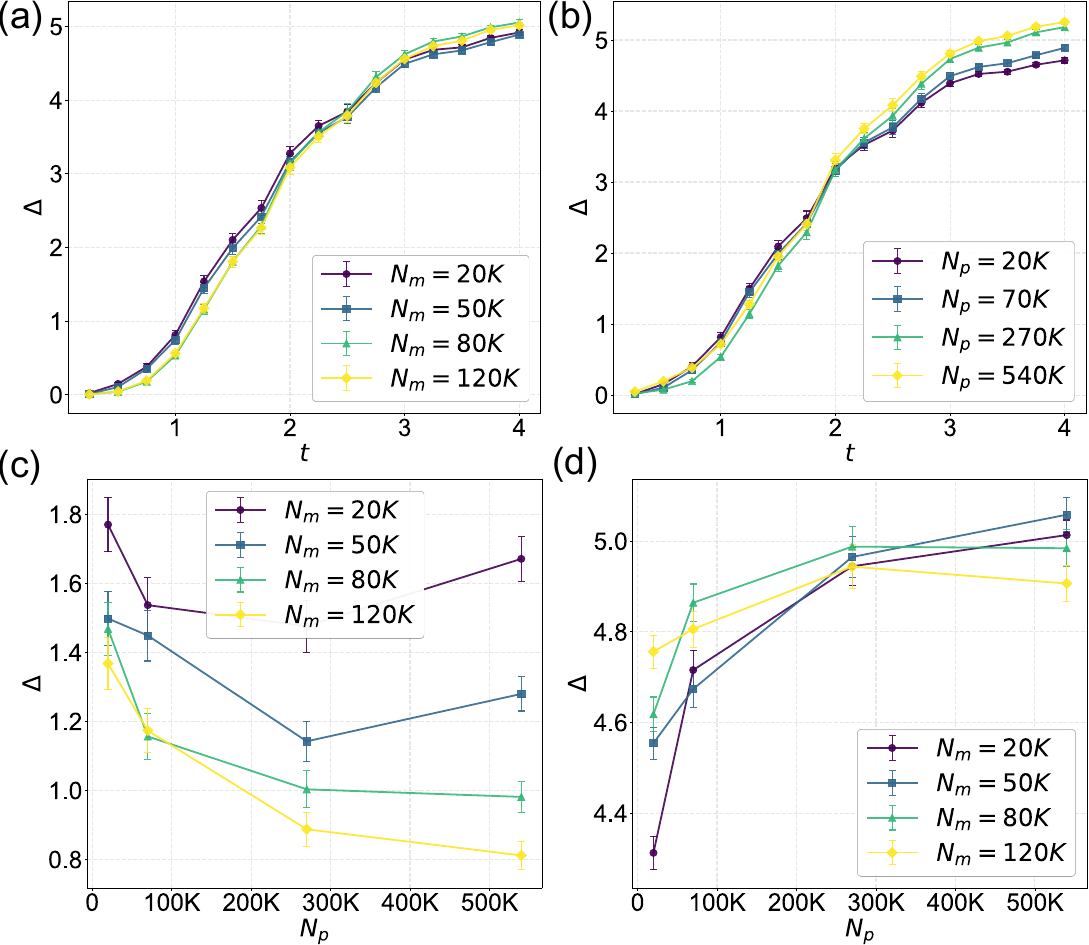}
\end{center}
\caption{Learnability transition in one-dimensional all-to-all circuits with $\mathcal{L}_{1}(\theta)$ as the loss function. Here $L=20$, $N_{e}=5 \times 10^3$ and $M=50$. (a) $\Delta$ as a function of $t$  for different $N_{m}$, with $N_{p} = 7 \times 10^4$ fixed. (b) $\Delta$ as a function of $t$ for different $N_{p}$, with $N_{m} = 5 \times 10^4$ fixed. (c,d) $\Delta$ as a function of $N_{m}$ and $N_{p}$ for $t=1.25$ and $t=3.5$, respectively.}
\label{supp_fig9}
\end{figure}

Repeating the resource analysis with $\mathcal{L}_{1}(\theta)$ reproduces the transition. As shown in Fig.~\ref{supp_fig9}, at small $t$ the uncertainty $\Delta$ falls as either $N_m$ or $N_p$ grows, while at large $t$ it saturates and eventually rises---the learnable and unlearnable phases as before. The transition is therefore not an artifact of the loss. One difference is that $\Delta$ can now exceed $2\ln(2)$: the model still outputs a nearly pure $\rho_{AB,m}$ at large $t$, but one nearly orthogonal to the true $\sigma_{AB,m}$, which inflates $\Delta$.

A related construction, considered in Ref.~\cite{hou2025}, instead uses the raw shadow state on $AB$,
\begin{equation}
\widetilde{\sigma}_{AB,m}^{s} = \left((U_{A}^{s})^{-1}\ket{m_{A}}\bra{m_{A}}U_{A}^{s}\right) \otimes \left((U_{B}^{s})^{-1}\ket{m_{B}}\bra{m_{B}}U_{B}^{s}\right),
\end{equation}
giving $\mathcal{L}_{2}(\theta) = -\mathbb{E}_{m}\mathbb{E}_{s}\text{Tr}(\rho_{AB,m}\widetilde{\sigma}_{AB,m}^{s})$. We show, however, that it is valid only when the post-measurement state $\sigma_{AB,m}$ has a flat entanglement spectrum.
Recall that the shadow snapshot is given by:
\begin{equation}
\sigma_{AB,m}^{s} = \left(3(U_{A}^{s})^{-1}\ket{m_{A}}\bra{m_{A}}U_{A}^{s}-I\right) \otimes \left(3(U_{B}^{s})^{-1}\ket{m_{B}}\bra{m_{B}}U_{B}^{s}-I\right),
\end{equation}
where $I$ denotes the $2\times 2$ identity matrix. Since $\mathbb{E}_{s}[\sigma_{AB,m}^{s}]=\sigma_{AB,m}$, we have:
\begin{equation}
\mathbb{E}_{s}[\widetilde{\sigma}_{AB,m}^{s}] = \frac{1}{9}(\sigma_{AB,m} + \sigma_{A,m}\otimes I + I \otimes \sigma_{B,m} + I \otimes I).
\end{equation}
Consequently, the loss function simplifies to:
\begin{equation}
\mathcal{L}_{2}(\theta) = -\frac{1}{9}\mathbb{E}_{m}\text{Tr}(\rho_{AB,m} \Omega),
\end{equation}
where we define the effective operator $\Omega \equiv \sigma_{AB,m} + \sigma_{A,m} \otimes I + I \otimes \sigma_{B,m} + I \otimes I$. We now show that for each measurement outcome $m$, the maximizer $\text{argmax}_{\rho_{AB,m}} \text{Tr}(\rho_{AB,m} \Omega)$ coincides with the true state $\sigma_{AB,m}$ if and only if $\sigma_{AB,m}$ has a flat entanglement spectrum. Let $\sigma_{AB,m}= \ket{\psi}\bra{\psi}$. Utilizing the Schmidt decomposition, we can write $|\psi\rangle = \sum_k \sqrt{\lambda_k} |k\rangle_A \otimes |k\rangle_B$. The reduced density matrices are:
\begin{equation}
\sigma_{A,m} = \sum_{k} \lambda_{k} \ket{k}_{A}\bra{k}_{A}, \quad \sigma_{B,m} = \sum_{k}\lambda_{k}\ket{k}_{B} \bra{k}_{B}.
\end{equation}
For $\text{argmax}_{\rho_{AB,m}} \text{Tr}(\rho_{AB,m} \Omega) = \sigma_{AB,m}$ to hold, the state $\ket{\psi}$ must be an eigenvector of $\Omega$, i.e.,  $\Omega \ket{\psi} \propto \ket{\psi}$. Applying $\Omega$ to the state $\ket{\psi}$ yields:
\begin{equation}
\Omega \ket{\psi} = \sum_{k} \sqrt{ \lambda_{k} }(2+ 2\lambda_{k}) \ket{k}_{A} \ket{k}_{B}.
\end{equation}
Therefore, the eigenvector condition requires the coefficient $2+2\lambda_{k}$ to be independent of $k$, which implies that all non-zero Schmidt coefficients $\lambda_{k}$ must be equal. Therefore, if the true post-measurement state $\sigma_{AB,m}$ has a nontrivial entanglement spectrum—as is typical in our setup—this loss function fails to correctly train the model.

To conclude, none of these four factors accounts for the loss of learnability, indicating that the learnability transition is physical rather than an artifact of the learning method.

\section{Experimental details and validation}
\label{sec:exp}

We now turn from the numerical analysis to the experimental validation of the transition, first through a noise-model simulation and then on a superconducting quantum processor. The noise simulation in Fig.~4(a,b) of the main text uses the Qiskit noise snapshot of $\text{ibm\_brisbane}$, which composes thermal relaxation from the calibrated $T_1,T_2$ and gate durations, depolarizing gate errors, and asymmetric readout; being snapshot-based and Markovian, it neglects drift and cross-talk. The hardware experiment in Fig.~4(c,d), performed on $\text{ibm\_marrakesh}$, establishes that the transition remains observable under present-day device noise; since it runs on the processor itself, this noise is the device's own rather than a model. Both the simulated and the on-device noise are characterized in Table~\ref{tab:backend}.

Because the protocol requires per-shot bitstrings, it is a sampling protocol, and the hardware data were taken with all sampling-compatible error mitigation: dynamical decoupling, Pauli and measurement twirling, and readout-calibrated (``robust'') classical shadows built from each qubit's measured readout error $p$. This calibration is needed because a symmetric readout bit-flip of probability $p$ attenuates each recorded $\pm1$ outcome by a factor $(1-2p)$, equivalently rescaling the traceless part of the single-qubit shadow channel by $(1-2p)$; inverting the ideal channel with the factor $3$ then biases each snapshot toward the maximally mixed state. Inverting the calibrated channel instead, i.e.\ replacing the single-qubit reconstruction $3\,U_{q}^{s\dagger}\ket{m_{q}}\bra{m_{q}}U_{q}^{s}-I$ by $\tfrac{3}{1-2p}\,U_{q}^{s\dagger}\ket{m_{q}}\bra{m_{q}}U_{q}^{s}-\tfrac12\big(\tfrac{3}{1-2p}-1\big)I$, whose identity offset is fixed by unit trace, removes this bias exactly and restores $\mathbb{E}_s[\sigma_{AB,m}^{s}]=\sigma_{AB,m}$. Because $\Delta$ and the training loss are linear in the snapshots, this per-shot de-biasing carries over directly to an unbiased $\Delta$; its only input is the independently measured readout error of each qubit, recorded here for the physical qubits carrying $A$ and $B$. It is worth emphasizing that zero-noise extrapolation, probabilistic error cancellation, and matrix-inversion readout mitigation are not applicable here: they return corrected expectation values or quasi-probabilities, whereas every stage of our pipeline operates on individual bitstrings.

The hardware run was constrained by the available quantum-processor time, which limited us to $M=5$ circuit realizations per depth, each evaluated on $N_{e}=5\times10^3$ shots. With so few realizations, a simple standard error over the circuits is unreliable. Therefore, we quantify the uncertainty of $\Delta$ with two more robust methods. The first is a hierarchical bootstrap. The per-shot entropy $S_{AB,m}^{\text{SC}}=-\text{Tr}(\sigma_{AB,m}^{s}\ln\rho_{AB,m})$ carries two nested sources of fluctuation---variation between circuit realizations and shadow shot noise within each realization---so for each of $B=2000$ bootstrap replicates we draw $M$ realizations with replacement and, within each, $N_{e}$ shots with replacement, and recompute $\Delta$ as the grand mean; the $95\%$ confidence intervals in Fig.~4(d) and Table~\ref{tab:hwci} are the $2.5$th and $97.5$th percentiles of the replicates, which assume neither Gaussian statistics nor large $M$. The second is a median-of-means estimator for the heavy-tailed $S_{AB,m}^{\text{SC}}$: we randomly partition the $M N_{e}$ per-shot values into $G=50$ groups, average within each, and take the median of the group means, which suppresses the influence of rare large snapshots. As listed in Table~\ref{tab:hwci}, the mean and median-of-means agree closely at every depth, and the shallow and deep regimes have non-overlapping confidence intervals, so a signature of the learnability transition is visible even at $M=5$.

\begin{table}[h]\centering
\caption{Device parameters for the numerical simulation and the hardware experiment. The $\text{ibm\_brisbane}$ column is the Qiskit snapshot underlying the noise model of the simulation [Fig.~4(a,b)]; the $\text{ibm\_marrakesh}$ column is the run-time calibration recorded at data acquisition for the hardware experiment [Fig.~4(c,d)]. Their native two-qubit gates are the echoed cross-resonance (ECR) and the controlled-$Z$ (CZ), respectively. Each entry is the device-wide median, and the bracketed numbers give the min--max range over all qubits or qubit pairs. These ranges include outlier and dead qubits, whereas the routed 20-qubit chain used low-error qubits, with the physical qubits carrying $A,B$ having readout errors of $0.3$--$0.5\%$.}\label{tab:backend}
\renewcommand{\arraystretch}{1.2}
\begin{tabular*}{4.8in}
{@{\extracolsep{\fill}}lcc}
\hline\hline
 & $\text{ibm\_brisbane}$ [Fig.~4(a,b)] & $\text{ibm\_marrakesh}$ [Fig.~4(c,d)] \\
\hline
\# qubits & 127 & 156 \\
two-qubit gate & ECR & CZ \\
$T_1$ ($\mu$s) & 232 [9.9--425] & 194 [6.6--521] \\
$T_2$ ($\mu$s) & 150 [17--412] & 102 [6.4--517] \\
1q ($\sqrt{X}$) error & $2.4\times10^{-4}$ & $3.3\times10^{-4}$ \\
2q gate error & $0.77\%$ & $0.29\%$ \\
readout error & $2.0\%$ [0.56--24] & $1.3\%$ [0.15--50] \\
\hline\hline
\end{tabular*}
\end{table}

\begin{table}[h]\centering
\caption{Uncertainty $\Delta$ versus depth $t$ for the hardware experiment ($\text{ibm\_marrakesh}$, $L=20$, $M=5$). Each row lists the mean $\Delta$, the median-of-means (MoM), and the $95\%$ hierarchical-bootstrap confidence interval. }\label{tab:hwci}
\renewcommand{\arraystretch}{1.1}
\begin{tabular*}{3.8in}
{@{\extracolsep{\fill}}ccccc}
\hline\hline
$t$ & $\Delta$ & MoM & CI low & CI high \\
\hline
$0.25$ & 0.170 & 0.158 & 0.063 & 0.274 \\
$0.50$ & 0.395 & 0.412 & 0.221 & 0.579 \\
$0.75$ & 0.477 & 0.510 & 0.377 & 0.579 \\
$1.00$ & 0.812 & 0.766 & 0.572 & 1.040 \\
$1.25$ & 0.917 & 0.905 & 0.726 & 1.087 \\
$1.50$ & 1.198 & 1.193 & 1.085 & 1.302 \\
$1.75$ & 1.283 & 1.296 & 1.238 & 1.322 \\
$2.00$ & 1.254 & 1.255 & 1.174 & 1.319 \\
$2.25$ & 1.286 & 1.287 & 1.174 & 1.371 \\
$2.50$ & 1.381 & 1.379 & 1.341 & 1.417 \\
$2.75$ & 1.360 & 1.368 & 1.322 & 1.392 \\
$3.00$ & 1.400 & 1.391 & 1.382 & 1.418 \\
$3.25$ & 1.391 & 1.390 & 1.372 & 1.411 \\
$3.50$ & 1.388 & 1.388 & 1.376 & 1.400 \\
$3.75$ & 1.400 & 1.396 & 1.385 & 1.417 \\
$4.00$ & 1.398 & 1.395 & 1.387 & 1.409 \\
\hline\hline
\end{tabular*}
\end{table}

\section{Additional numerical results for one-dimensional circuits}

\begin{figure}[htbp]
\begin{center}
\includegraphics[width=0.9\textwidth, clip=true]{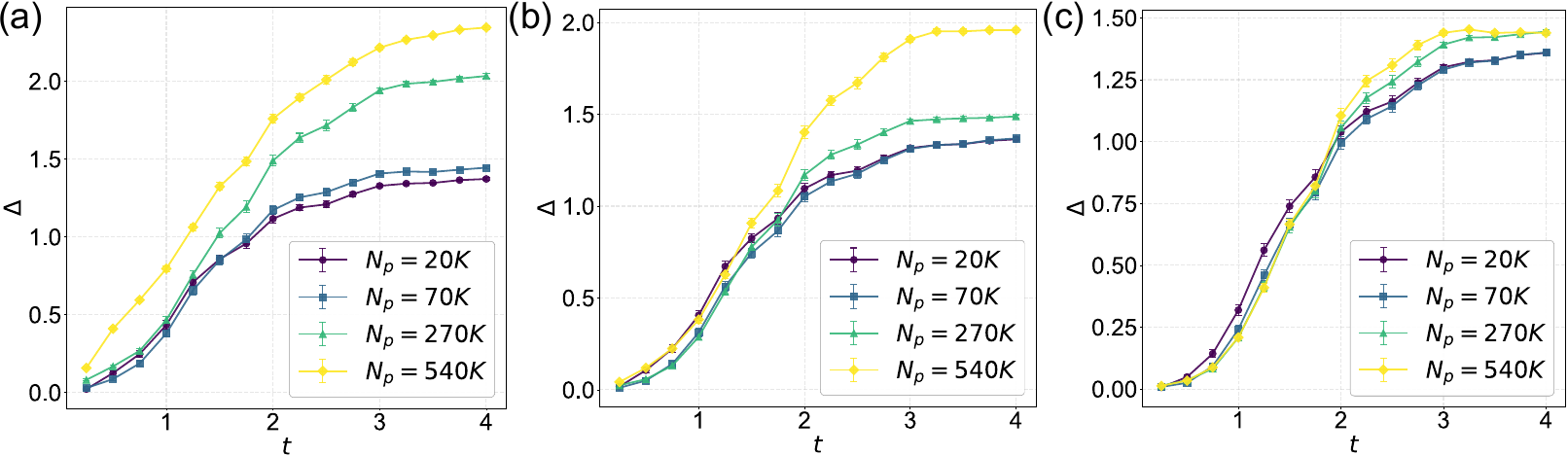}
\end{center}
\caption{Learnability transition in one-dimensional all-to-all circuits. $\Delta$ as a function of $t$ for different $N_{p}$, with different $N_{m}$ being fixed. Here $L=20$, $N_{e}=5 \times 10^3$ and $M=50$. (a) $N_{m}=2 \times 10^4$. (b) $N_{m}=5 \times 10^4$. (c) $N_{m}=1.2 \times 10^5$. }
\label{supp_fig10}
\end{figure}

We provide further numerical evidence for the learnability transition in one-dimensional all-to-all circuits. In Fig.~\ref{supp_fig10}, we show $\Delta$ for a range of combinations of $N_{m}$ and $N_{p}$ across all circuit depths $t$. A clear trend is that as $N_{m}$ increases, overfitting effects are mitigated, and  a pronounced crossing becomes visible—below the crossing, larger $N_{p}$ reduces $\Delta$; whereas above it, increasing $N_{p}$ leads to larger $\Delta$. This behavior is consistent with the presence of two distinct phases, corresponding to the learnable and unlearnable regimes.

\section{Two-dimensional nearest-neighbor circuits}

\begin{figure}[tbp]
\begin{center}
\includegraphics[width=0.9\textwidth, clip=true]{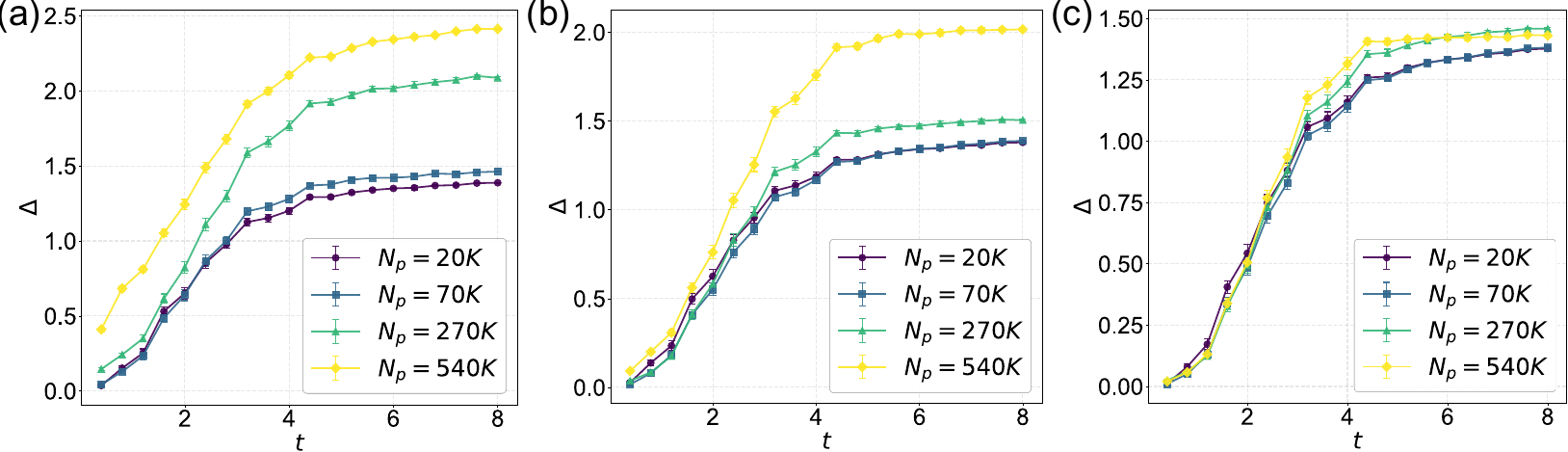}
\end{center}
\caption{Learnability transition in two-dimensional nearest-neighbor circuits. $\Delta$ as a function of $t$ for different $N_{p}$, with different $N_{m}$ being fixed. Here $L=5 \times 5$, $N_{e}=5 \times 10^3$ and $M=50$. (a) $N_{m}=2 \times 10^4$. (b) $N_{m}=5 \times 10^4$. (c) $N_{m}=1.2 \times 10^5$. }
\label{supp_fig11}
\end{figure}

We finally ask whether the learnability transition is special to the all-to-all connectivity studied above, by turning to two-dimensional nearest-neighbor circuits. We consider $L=5\times5$ qubits on a square lattice with periodic boundary conditions, applying at each time step $L\delta t$ Haar-random two-qubit gates restricted to neighboring pairs. Although this geometry is spatially local, it too hosts a finite-depth computational transition~\cite{napp2022,lee2025,bene2025,mcginley2025,cao2025}, analogous to the teleportation transition~\cite{bao2024} underlying the all-to-all case, across which classical simulation of the shallow circuit changes from tractable to intractable. In Fig.~\ref{supp_fig11}, we show $\Delta$ for a range of combinations of $N_{m}$ and $N_{p}$ across all circuit depths $t$. As in one dimension, increasing $N_{m}$ mitigates overfitting and reveals a pronounced crossing—below the crossing, larger $N_{p}$ reduces $\Delta$; whereas above it, increasing $N_{p}$ leads to larger $\Delta$.

\begin{figure}[htbp]
\begin{center}
\includegraphics[width=0.65\textwidth, clip=true]{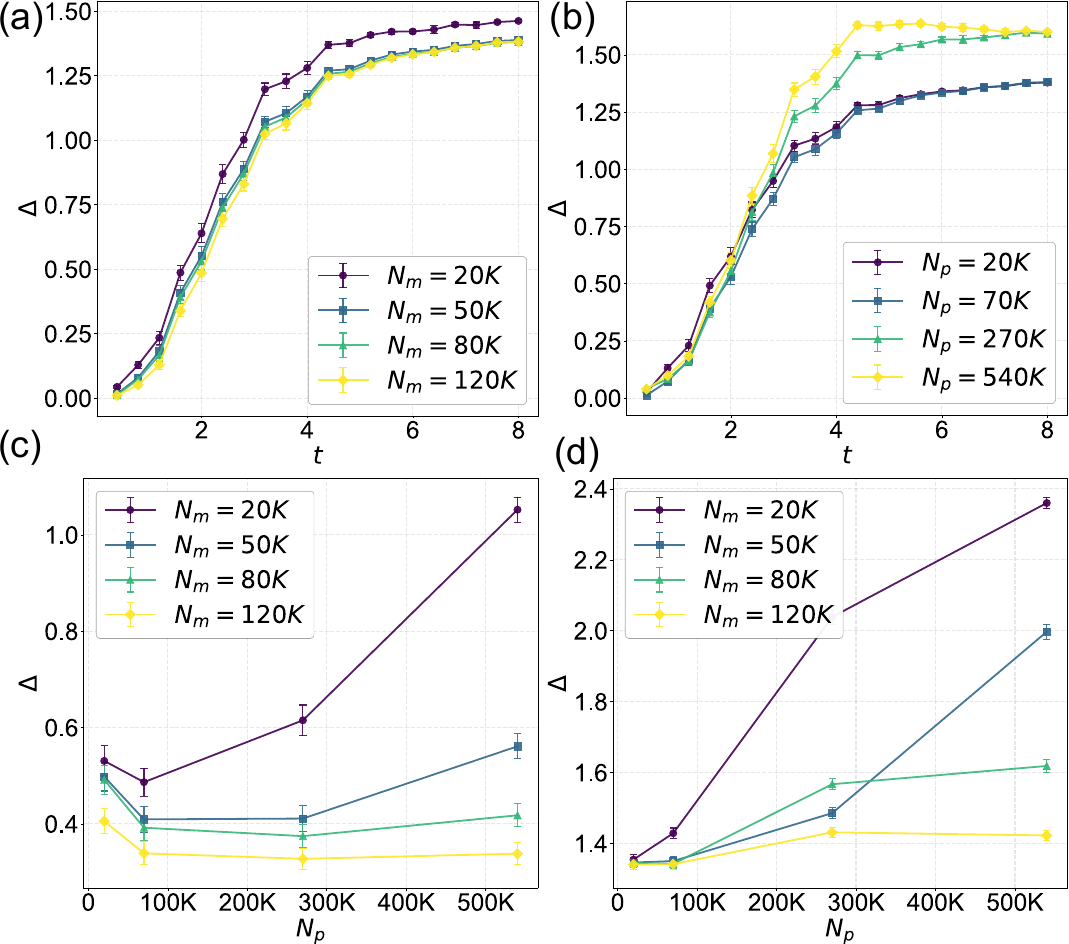}
\end{center}
\caption{Learnability transition in two-dimensional nearest-neighbor circuits for $L=5\times5$. (a) $\Delta$ as a function of $t$ for varying $N_{m}$, with $N_{p}=7\times10^4$ fixed. (b) $\Delta$ as a function of $t$ for varying $N_{p}$, with $N_{m}=8\times10^4$ fixed. (c,d) $\Delta$ as a function of $N_{m}$ and $N_{p}$ at representative depths $t=1.6$ and $t=6.4$, respectively. Here $N_{e}=5\times10^3$ and $M=50$.}
\label{supp_fig12}
\end{figure}

We repeat the resource analysis of the main text for this geometry. Fig.~\ref{supp_fig12}(a) and (b) show that at shallow depth $\Delta$ decreases as either the measurement budget $N_m$ or the number of parameters $N_p$ is increased, whereas at large depth it saturates and eventually grows through overfitting. Fig.~\ref{supp_fig12}(c) and (d) show the combined dependence on $N_m$ and $N_p$ at $t=1.6$ and $t=6.4$, which cleanly separates a learnable depth from an unlearnable one, mirroring the one-dimensional case. The learnable and unlearnable phases thus appear in two dimensions as well, consistent with the finite-depth computational transition of this geometry, so the learnability transition is not an artifact of the all-to-all connectivity but a generic feature of monitored circuit dynamics. An order parameter analogous to $\varphi_L$ of Sec.~\ref{sec:scaling} could be constructed here as well, but resolving its finite-size crossing would require exact classical simulation of several larger two-dimensional lattices, which is computationally prohibitive; we therefore restrict the two-dimensional study to this qualitative comparison.

\clearpage